\documentclass[%
reprint,
nofootinbib,
superscriptaddress,
amsmath,
amssymb,
aps,
prb
]{revtex4-2}
\usepackage[%
colorlinks=true,
urlcolor=blue,
linkcolor=blue,
citecolor=blue
]{hyperref}

\usepackage{graphicx}

\begin{document}

\title{Competing orders in the honeycomb lattice $t$-$J$ model}
\author{Zheng-Tao Xu}
\affiliation{State Key Laboratory of Low-Dimensional Quantum and Department of Physics, Tsinghua University, Beijing 100084, China}
\author{Zheng-Cheng Gu}
\email{zcgu@phy.cuhk.edu.hk}
\affiliation{Department of Physics, The Chinese University of Hong Kong, Shatin, New Territories, Hong Kong, China}
\author{Shuo Yang}
\email{shuoyang@tsinghua.edu.cn}
\affiliation{State Key Laboratory of Low-Dimensional Quantum and Department of Physics, Tsinghua University, Beijing 100084, China}
\affiliation{Frontier Science Center for Quantum Information, Beijing 100084, China}
\affiliation{Hefei National Laboratory, Hefei 230088, China}

\begin{abstract}
We study the honeycomb lattice $t$-$J$ model using the fermionic tensor network approach.
By examining the ansatz with various unit cells, we discover several different stripe states with different periods that compete strongly with uniform states.
At very small doping $\delta < 0.05$, we find almost degenerate uniform $d$-wave superconducting ground states coexisting with antiferromagnetic order.
While at larger doping $\delta > 0.05$, the ground state is an approximately half-filled stripe-ordered state, where the stripe period decreases with increasing hole doping $\delta$.
Furthermore, the stripe states with the lowest variational energy always display $d_{x^2-y^2}$-wave pairing symmetry. 
The similarity between our results and those on the square lattice contributes to a more comprehensive understanding of doped Mott insulators.
\end{abstract}

\maketitle

\section{Introduction}
The $t$-$J$ model, which can be derived from the strong-coupling limit of the Hubbard model, is one of the simplest and most important models for strongly correlated systems. 
The Hamiltonian of the $t$-$J$ model, describing a doped Mott insulator, reads
\begin{equation}
H = -t\sum_{\langle ij\rangle , \sigma}\left( \tilde{c}_{i,\sigma}^\dagger \tilde{c}_{j,\sigma} + h.c.\right) + J\sum_{\langle i, j\rangle} \left( \vec{S}_i\cdot\vec{S}_j - \frac{1}{4}\hat{n}_i \hat{n}_j\right), \nonumber
\end{equation}
where $\tilde{c}_{i,\sigma} = \hat{c}_{i\sigma}(1-\hat{n}_{i\bar{\sigma}})$ is the electron operator defined in the no-double-occupancy subspace. 
In the past three decades, it has been suggested that such a simple model on a square lattice could capture the fundamental properties of high-$T_c$ cuprates.
Many cutting-edge methods \cite{RevModPhys.78.17,Zheng1155,PhysRevB.84.041108,PhysRevLett.113.046402,PhysRevB.98.205132,PhysRevB.97.045138,PhysRevB.93.035126,Jiang1424,PhysRevB.100.195141,doi:10.1146/annurev-conmatphys-090921-033948} have produced fascinating results on the competing order nature in the $t$-$J$ and Hubbard models on the square lattice. 
For example, the ground state is a period 8 stripe state without $d$-wave superconducting at $\delta=1/8$ \cite{Zheng1155}, whereas it is a period 4 stripe state with $d$-wave superconducting by adding next-nearest neighbor hoppings around $t'=-0.25t$ \cite{Jiang1424,PhysRevB.100.195141}. 
    
Recently, the Hubbard model on a honeycomb lattice has also been studied intensively since the discovery of superconductivity in twisted bilayer graphene.
In contrast to the square lattice case, the Hubbard model on the honeycomb lattice has a metal-insulator transition (MIT) with a critical $U_c$ around $3.8$ \cite{PhysRevX.3.031010, PhysRevX.6.011029}.
In the weak-coupling limit, most previous results \cite{PhysRevB.75.134512,PhysRevB.81.085431,PhysRevB.84.121410,PhysRevB.97.075127,PhysRevB.90.054521,PhysRevX.4.031040,PhysRevB.85.035414,PhysRevB.86.020507,Nandkishore2012,PhysRevB.94.115105,BS2014} suggest that uniform $d+id$-wave superconductivity may occur in doped graphene systems using various approaches, such as a mean field theory \cite{PhysRevB.75.134512}, quantum Monte Carlo (QMC) \cite{PhysRevB.81.085431,PhysRevB.84.121410,PhysRevB.97.075127,PhysRevB.90.054521,PhysRevX.4.031040,PhysRevB.85.035414}, renormalization group (RG)  \cite{PhysRevB.85.035414,PhysRevB.86.020507,Nandkishore2012}, and dynamical cluster approximation (DCA) \cite{PhysRevB.94.115105}.
In the strong-coupling limit, $d+id$ superconductivity is also discovered in the $t$-$J$ model using the Grassmann tensor product state (GTPS) method \cite{PhysRevB.88.155112}.
Other possible competing orders including $s$-wave, even $f$-wave pairing symmetries \cite{PhysRevB.86.020507,PhysRevB.94.115105,PhysRevB.92.085121} as well as $p+ip$-wave pairing symmetry \cite{PhysRevLett.98.146801, PhysRevB.92.085121, PhysRevB.101.205147} have also been discovered in doped graphene systems or the infinite-$U$ Hubbard model using GTPS \cite{PhysRevB.101.205147}.
Very recently, stripe order has been found in the Hubbard model at $1/16$ and $1/12$ dopings with $U=8$ using the density matrix renormalization group (DMRG) and the auxiliary-field quantum Monte Carlo (AFQMC) method \cite{PhysRevB.103.155110, PhysRevB.105.035111}. 

\begin{figure}[tbp]
\centering
\includegraphics[width=1.0\columnwidth]{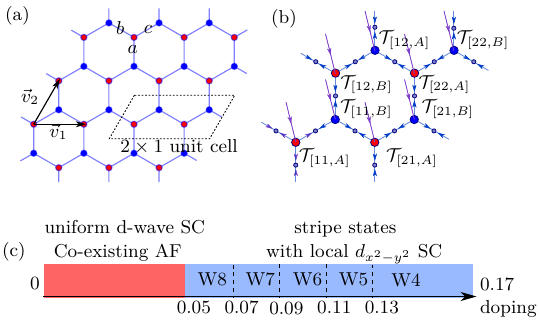} 
\caption{
(a) The honeycomb lattice in the thermodynamic limit. The blue (red) dot represents the $A$ ($B$) site of the two sub-lattices. 
The dashed square indicates the $L_1\times L_2 = 2\times 1 $ unit cell, and the primitive vectors are represented by the two arrows $\vec{v}_1$ and $\vec{v}_2$.
The lattice has three different directions: $a$, $b$, and $c$.
(b) The fermionic tensor network state in the $\mathbb{Z}_2$-graded formalism.
The small dots represent the Schmidt weights $\Lambda$ on the bonds between neighboring sites.
(c) Phase diagram for $t/J=3.0$ and the $L_1\times 1$ cell as a function of hole doping $\delta$ from $0.0$ to $0.17$.
Here W$l_x$ means the period of the stripe state is $l_x$.
}
\label{fig:1}
\end{figure}

\begin{figure}[tbp]
\centering
\includegraphics[width=0.95\columnwidth]{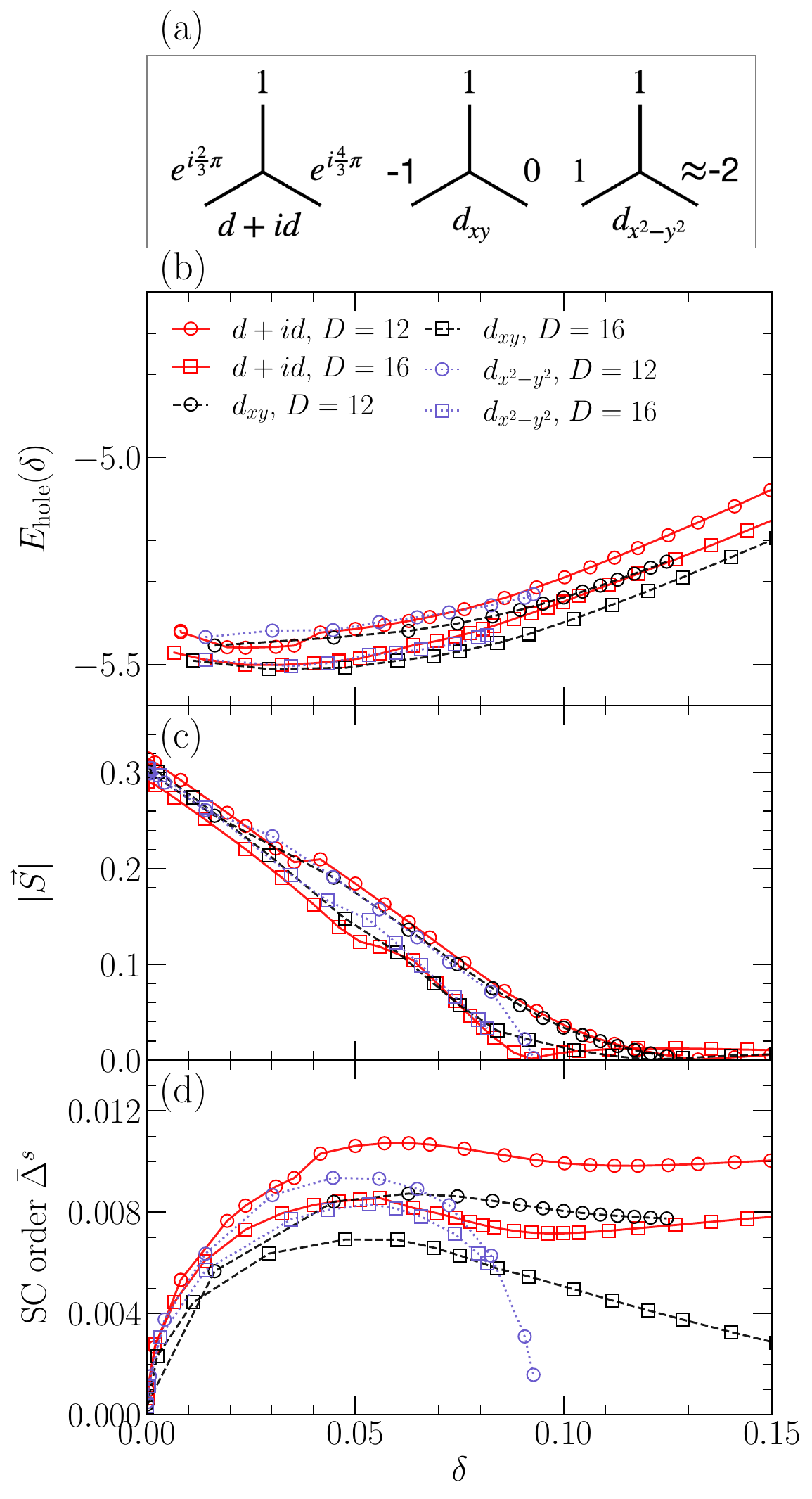} 
\caption{
(a) Three uniform states with distinct $d$-wave pairing symmetries.
(b)-(d) Physical quantities for uniform states as a function of doping $\delta$ with $J/t=3.0$ and different bond dimensions $D=12,16$.
(b) The energy per hole $E_{{\rm hole}}(\delta) = [E_0(\delta) - E_0(0)]/\delta$.
(c) The staggered magnetization $|\vec{S}|$.
(d) The average spin-singlet pairing SC order of three nearest-neighbor bonds $\bar{\Delta}^s = \frac{1}{3}(|\Delta^s_a| + |\Delta^s_b| + |\Delta^s_c|)$.
}
\label{fig:2}
\end{figure} 

Experimentally, long-range antiferromagnetic (AFM) order has been found in several materials with honeycomb lattice structures, such as InCu$_{2/3}$V$_{1/3}$O$_3$ \cite{KATAEV2005310}. 
It suggests that Cu spins in the material can be modeled as a honeycomb spin-$1/2$ Heisenberg model with a strong AFM coupling $J\sim 140$K.

In this paper, we use fermionic tensor network states \cite{gu2010grassmann,PhysRevB.88.115139,PhysRevB.95.075108,Bultinck_2017,PhysRevB.80.165129,PhysRevA.81.010303,PhysRevB.81.165104,PhysRevA.81.052338} to investigate the global phase diagram and competing orders in the honeycomb $t$-$J$ model with $t/J=3.0$ in the thermodynamic limit. 
The phase diagram in Fig. \ref{fig:1}(c) summarizes the results.
Due to nearly degenerated ground states, we obtain various states with different superconducting pairing symmetries on the bipartite unit cell and the $L_1\times 1 $ ($2 \leq L_1 \leq 8$) supercells at hole doping $\delta < 0.2$.
While previous research \cite{PhysRevB.88.155112} suggests a $d+id$-wave superconducting ground state breaking time-reversal symmetry, we find two more uniform superconducting states on the bipartite unit cell with $d_{xy}$-wave and $d_{x^2-y^2}$-wave pairing symmetry.
Interestingly, all three uniform superconducting states with different pairing symmetries are almost degenerate at low doping $\delta < 0.05$. 
We further discover nonuniform states with stripe orders on $L_1\times 1$ ($L_1\neq 1$) supercells that locally show $d_{xy}$ or $d_{x^2-y^2}$-wave pairing symmetry. 
At larger doping $\delta > 0.05$, these stripe states always have lower energy than the uniform states. 
Compared with $d_{xy}$-wave stripe states, $d_{x^2-y^2}$-wave stripe states are favored at $\delta < 0.17$. 
The stripe period decreases with increasing $\delta$, which is very similar to previous studies on the square lattice \cite{PhysRevB.84.041108,PhysRevLett.113.046402,Zheng1155,PhysRevB.93.035126,PhysRevB.98.205132,PhysRevB.97.045138}. 
Remarkably, the lowest energy stripe states are nearly half-filled with $\rho_l\equiv \delta \times L_x \sim 0.55$.

\section{Method}
The ground-state wave function is obtained via the simple update (SU) scheme based on the imaginary time evolution technique \cite{PhysRevLett.98.070201}.
We choose a modest $\Delta\tau$ that decreases gradually to ensure convergence and efficiency.
The change in the average Schmidt weight at the end of SU is less than $10^{-9}$.
After that, the 2D tensor network is contracted using the variational uniform matrix product state (VUMPS) method \cite{10.21468/SciPostPhysLectNotes.7,PhysRevB.98.235148,PhysRevB.97.045145}.
The relative errors for physical quantities such as ground-state energy, magnetization, and superconductivity are in the order of $10^{-4}$ for an appropriate environment bond dimension $\chi$.
We examine both uniform states with a bipartite unit cell and nonuniform states with $L_1\times 1$ ($L_1\neq 1$) supercells.

\begin{figure*}[tbp]
\centering
\includegraphics[width=2.0\columnwidth]{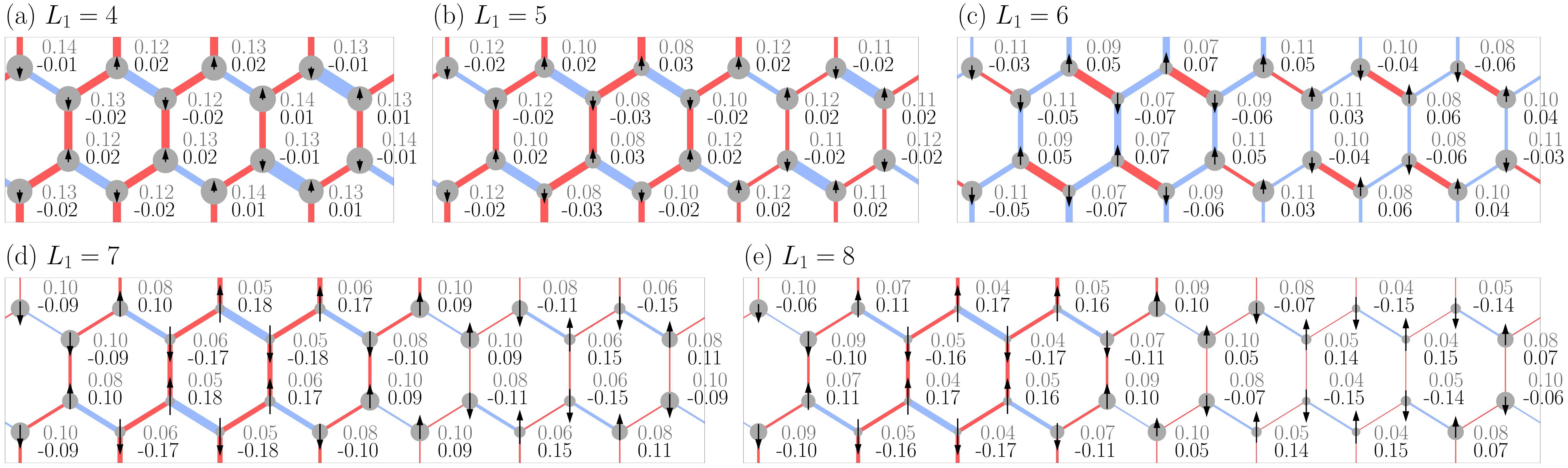} 
\caption{
Patterns of $d_{x^2-y^2}$-wave stripe states with different widths and $D=14$.
(a) W4 state at $\delta = 0.13$.
(b) W5 state at $\delta = 0.11$.
(c) W6 state at $\delta = 0.095$.
(d) W7 state at $\delta = 0.075$.
(e) W8 state at $\delta = 0.065$.
The diameter of each disk scales with local hole doping, with the value marked by grey.
The length of each arrow scales with local magnetization, with the value marked by black.
The bond between two sites represents the local spin-singlet pairing SC order with a positive (red/dark grey) or negative (blue/light grey) sign, and the width scales with pairing amplitude.
SC orders along the $a$/$c$ direction and the $b$ direction have opposite signs for $d_{x^2-y^2}$-wave stripe states.
The favored stripe states display a $\pi$-phase shift in the AFM order.
}
\label{fig:3}
\end{figure*}

\begin{figure}[tbp]
\centering
\includegraphics[width=0.95\columnwidth]{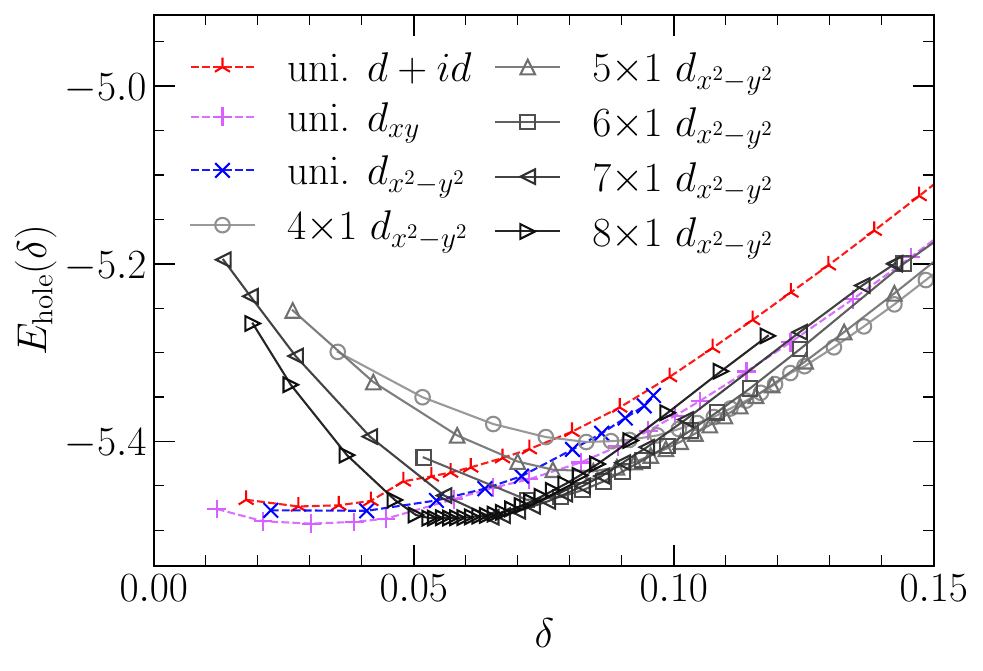} 
\caption{
Energies per hole of $d_{x^2-y^2}$-wave stripe states and uniform states for $t/J=3.0$ and $D = 14$. 
There is a transition from uniform states to stripe stats at $\delta_c \approx 0.05$.
}
\label{fig:4}
\end{figure}

\begin{figure}[tbp]
\centering
\includegraphics[width=1.0\columnwidth]{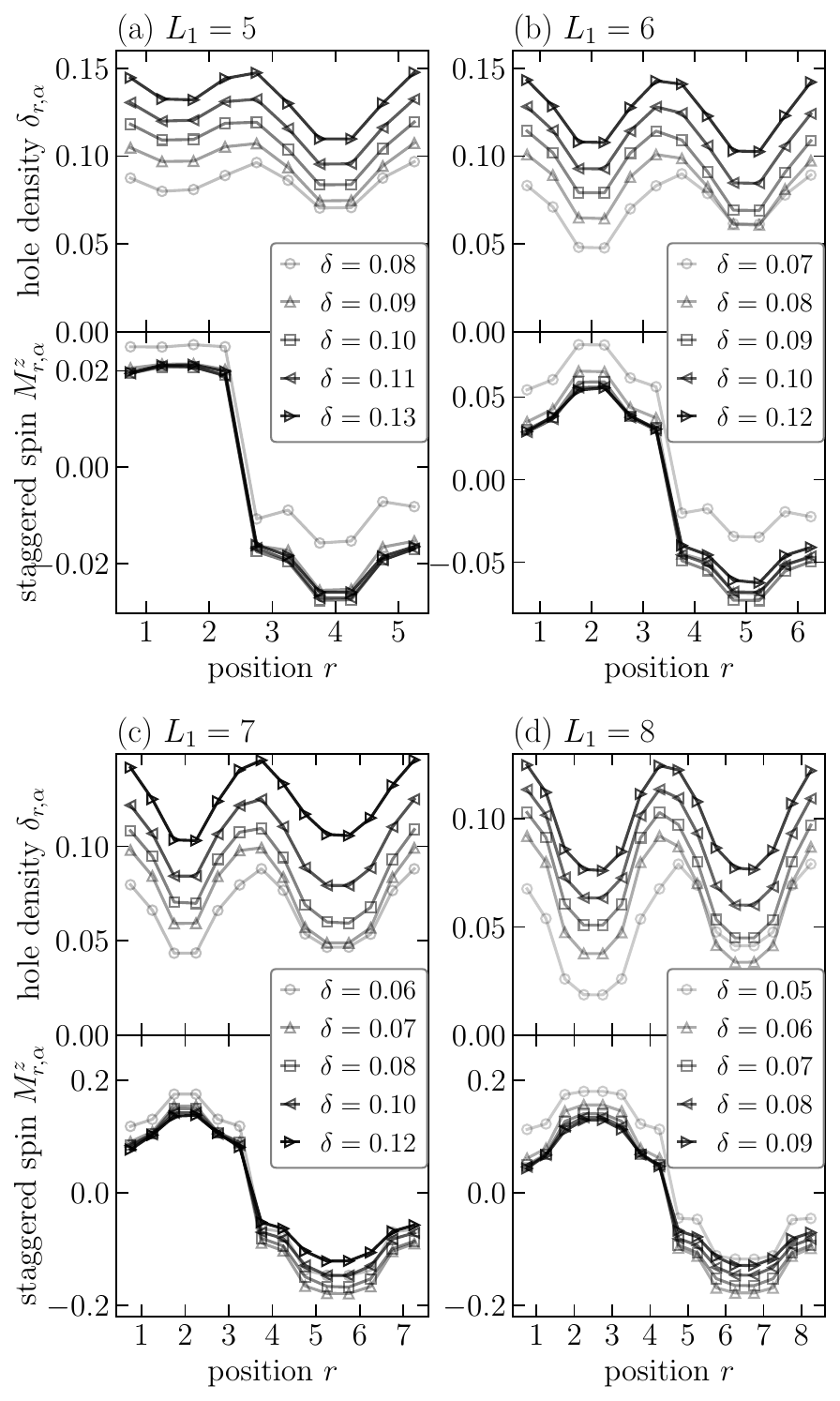} 
\caption{
Results for $d_{x^2-y^2}$-wave stripe states with $D=14$, $L_1=5$, $6$, $7$, $8$, and different hole doping $\delta$.
The top (bottom) figure for each $L_1$ depicts the local hole density $\delta_{r,\alpha}$ (staggered magnetic moment $M^z_{r,\alpha}$) as a function of sub-lattice position along the $\vec{v}_1$ direction.
}
\label{fig:5}
\end{figure}

\begin{figure}[btp]
\centering
\includegraphics[width=0.95\columnwidth]{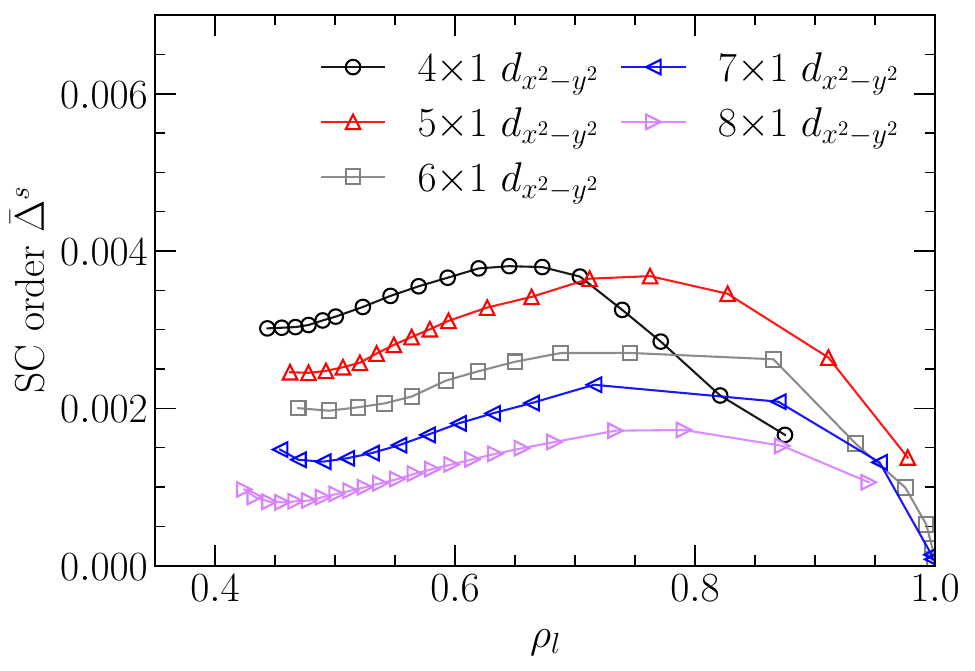} 
\caption{
The average pairing amplitudes of $d_{x^2-y^2}$-wave stripe states are plotted against hole density per unit length $\rho_l = \delta \times L_1$ with $D = 14$.
}
\label{fig:6}
\end{figure}

\section{Results}

\subsection{Uniform States}
We first investigate the uniform ansatz with only two different tensors in each unit cell. 
A chemical potential term is introduced to control hole doping $\delta$.
In the thermodynamic limit, the charge $U(1)$ symmetry would be broken spontaneously, such that the superconducting (SC) order can be detected directly in the spin-singlet channel in real space, e.g., $\hat{\Delta}^s_{ij} = \frac{1}{\sqrt{2}} \langle  \hat{c}_{i\uparrow}\hat{c}_{j\downarrow} - \hat{c}_{i\downarrow}\hat{c}_{j\uparrow}\rangle$.
The $d+id$-wave state with $\vec{\Delta}^s\sim(\Delta_a^s=1,\Delta_b^s=e^{i\frac{2}{3}\pi},\Delta_c^s=e^{i\frac{4}{3}\pi})$ has been previously discovered in Ref. \cite{PhysRevB.88.155112}. 
Interestingly, we find two more states with $E_{2g}$ symmetry: the $d_{xy}$-wave state with $\vec{\Delta}^s\sim (1,-1,0)$ and the $d_{x^2-y^2}$-wave state with $\vec{\Delta}^s\sim(1,1,\approx -2)$ in Fig. \ref{fig:2}(a).
In the following, we will discuss the variational energy, magnetism, and superconductivity for these uniform states.

Figure \ref{fig:2}(b) shows the energy comparison for three competing uniform states as a function of hole doping $\delta$.
Here we plot the energy per hole $E_{{\rm hole}}(\delta) = [E_0(\delta) - E_0(0)]/\delta$, where $E_0(0) = -0.91955$ at $\delta = 0$ is from a QMC calculation for the Heisenberg model on the honeycomb lattice \cite{PhysRevB.83.134421}.
We display energies for these states with various bond dimensions $D=12,16$. 
We further plot the $1/\chi$ dependence of ground-state energy in Appendix \ref{sec:MRandD}, which indicates that the relative error is smaller than the symbol size in Fig. \ref{fig:2}(b).
At small hole doping $\delta < 0.05$, the energies for three uniform states are incredibly close to each other, implying that the $d+id$-wave state may be a complex combination of the $d_{xy}$ and $d_{x^2-y^2}$-wave state.
When hole doping is increased with $\delta > 0.06$, the energy of the $d_{xy}$-wave state is slightly lower than that of the $d+id$-wave state for each $D$.
However, such a slight difference might be due to the explicit $C_3$ symmetry breaking of the VUMPS algorithm. 
For comparison, we also apply the Grassmann tensor renormalization group (GTRG) approach \cite{PhysRevLett.99.120601} to contract the two-dimensional tensor-network state (see Appendix \ref{sec:MRandD} for details) and find approximate degenerate energies for all three uniform states.

Figure \ref{fig:2}(c) depicts staggered magnetization $|\vec{S}|$ as a function of hole doping $\delta$.
The staggered magnetizations for three uniform states are found to be highly comparable and have a similar trend.
The antiferromagnetic (AF) order decreases (approximately) linearly with increasing hole doping and vanishes at $\delta \gtrsim 0.1$.
This is consistent with previous studies for the honeycomb lattice $t$-$J$ model \cite{PhysRevB.88.155112}.
In the extrapolation with $1/D$, we observe that the AFM order vanishes at $\delta \gtrsim 0.07$ for all three uniform states in Appendix \ref{sec:MRandD}.

The average SC order parameter $\bar{\Delta}^s = (|\Delta^s_a| + |\Delta^s_b| + |\Delta^s_c|)/3$ is shown in Fig. \ref{fig:2}(d) for three uniform states, where $a$, $b$, and $c$ indicate three directions.
The SC order of the $d_{x^2-y^2}$-wave state falls rapidly at $\delta \approx 0.1$ for each $D$.
The SC order of the $d_{xy}$-wave state exists when $\delta > 0.15$ for each $D$, but it decreases rapidly as $D$ increases.
Unlike the other two uniform states, the SC order of the $d+id$-wave can exist at large hole doping, even at $\delta > 0.15$, which is consistent with previous findings.
We also find that the SC order coexists with the AFM order for all uniform states at hole doping $\delta < 0.09$.

\subsection{The $d_{x^2-y^2}$-wave Stripe States}
We discover stripe states with different periods by investigating nonuniform ansatz with $L_1\times 1$ $(L_1 \geq 2)$ supercells.
The size of a unit cell determines the stripe period, and we call the W5 stripe state if the unit cell is a $5\times 1$ honeycomb lattice.
Two kinds of stripe states with $d_{xy}$ or $d_{x^2-y^2}$-wave pairing symmetry have lower energy than the uniform states at larger hole doping.
Moreover, the stripe states with $d_{x^2-y^2}$-wave pairing symmetry have lower energy than the $d_{xy}$-wave stripe states.
Therefore, we focus on $d_{x^2-y^2}$-wave stripe states in the main text and discuss more details for $d_{xy}$-wave stripe states in Appendix \ref{sec:dxySS}.
As shown in Fig. \ref{fig:3}, we plot patterns of $d_{x^2-y^2}$-wave stripe states with various stripe periods.
Here the $d_{x^2-y^2}$-wave pairing symmetry is characterized by positive SC orders along the $a$ and $c$ directions and negative SC orders along the $b$ direction, or vice versa.

In Fig. \ref{fig:4}, we compare energies per hole between uniform and stripe states as a function of hole doping $\delta$.
At $\delta_c \approx 0.05$, there is a phase transition from uniform $d$-wave states to stripe states.
When $\delta > 0.05$, stripe states with periods ranging from $4$ to $8$ strongly compete, and the preferred stripe period gradually decreases as hole doping $\delta$ increases.
The shift of stripe period as a function of $\delta$ has also been found in the ground-state phase diagram of the square Hubbard model with $U/t=10.0$ \cite{PhysRevB.97.045138, PhysRevB.98.205132}, and there is a similar shift of stripe period as a function of $t'/t$ in the square $t$-$t'$-$U$ model \cite{PhysRevB.100.195141}.
In addition, the $D = 10$ and $D = 12$ ansatz exhibit a similar trend at phase transitions among uniform states and stripe states with different stripe periods. 
Therefore, we believe $D = 14$ is large enough to determine the global phase diagram. 

The staggered spin density and hole density for these stripe states with $L_1 = 5$, $6$, $7$, and $8$ are depicted in Fig. \ref{fig:5}.
For each $L_1$, the top figure shows the local hole doping $\delta_{r,\alpha} = 1 - \langle \hat{n}_{r,\alpha,\uparrow}\rangle - \langle \hat{n}_{r,\alpha,\downarrow}\rangle$, while the bottom figure shows the local staggered magnetization $M^z_{r,\alpha} = (-1)^{\alpha} \langle S^z_{r,\alpha}\rangle$.
Here, $r$ denotes the position of the unit cell in the $\vec{v}_1$ direction, and $\alpha=0$ or $1$ denotes the $A$ or $B$ sub-lattice.
The staggered magnetization undergoes a $\pi$-phase shift along the $\vec{v}_1$ direction when $L_1 \geq 4$.
The period of spin density wave (SDW) is approximately twice as large as the period of charge density wave (CDW), similar to the results in the square \cite{Zheng1155, PhysRevB.97.045138, PhysRevB.98.205132} and honeycomb Hubbard models \cite{PhysRevB.105.035111, PhysRevB.103.155110}.
Furthermore, we notice that at small hole doping close to half filling $\rho_l \lesssim 0.4$, these stripe states exhibit AFM order without a $\pi$-phase shift of the staggered magnetization but have higher energy than uniform states.

Finally, we study the spin-singlet pairing SC order of the favored stripe states with $\pi$-phase shift.
Figure \ref{fig:6} shows the average SC order $\bar{\Delta}^s = \sum_{i,j} (|\Delta_{ija}^s| + |\Delta_{ijb}^s| + |\Delta_{ijc}^s|) /(3L_1L_2)$ as a function of hole density per unit cell $\rho_l = \delta \times L_1$, where $\Delta_{ij\alpha}$ indicates the SC order on the bond along the $\alpha$ direction ($\alpha=a,b,c$) from the site $[ij,A]$.
Local maximums of SC orders are around $\rho_l \approx 0.75$ for stripe states with different stripe periods, similar to the result in the square $t$-$J$ model \cite{PhysRevB.84.041108}.
In addition, the pairing amplitude decreases as the stripe period increases, indicating the competition between the stripe order and the SC order. 
However, the pairing amplitude of stripe states on the honeycomb lattice is weak ($\sim 10^{-3}$), whereas the pairing amplitude of the W5 stripe state in the square $t$-$J$ model reaches $10^{-2}$ at $\delta > 0.05$ using infinite projected entangled pair states (iPEPS) \cite{PhysRevLett.113.046402}.

\section{Summary and Conclusion}
We have investigated the ground-state properties of the $t$-$J$ model on the honeycomb lattice with $t/J=3.0$ using the fermionic tensor network approach.
We observe three nearly degenerate uniform states with different pairing symmetries ($d+id$-wave, $d_{xy}$-wave, and $d_{x^2-y^2}$-wave) at small hole doping, and stripe states with lower energy than uniform states at $\delta > 0.05$. 
For these stripe states, the stripe period decreases with increasing hole doping, and the period of charge density wave is half of the spin density wave. 
Furthermore, the superconductivity of stripe states is described as a local $d_{x^2-y^2}$ pairing symmetry.
For various stripe states, the SC order reaches a maximum at $\rho_l \approx 0.75$, but the SC order is weakened as the stripe period increases.
Compared to the amplitude of uniform states, the SC order is also suppressed by the stripe order.

Similar to the results in the square Hubbard model \cite{PhysRevB.84.041108, PhysRevLett.113.046402, Zheng1155, PhysRevB.97.045138, PhysRevB.98.205132}, the competing order nature between uniform superconductivity and nonuniform half-filled stripe states might play a significant role in the emergence of superconductivity in the strong coupling region. 
It would be of great importance to understand what kind of uniform states could be stabilized at low doping on the square lattice, and previous studies suggest a uniform $d$-wave state. 
The discovery of similarities between the honeycomb and the square $t$-$J$ model could pave the way for further research on superconductivity in honeycomb materials, which could help elucidate the underlying mechanism of high-$T_c$ SCs.

\section*{Acknowledgement}
We thank Mingpu Qin for providing some data in Ref. \cite{PhysRevB.105.035111} for comparison.
Z.T.X. and S.Y. are supported by the National Natural Science Foundation of China (NSFC) (Grant No. 12174214, No. 11804181, and No. 92065205), the National Key R\&D Program of China (Grant No. 2018YFA0306504), and the Innovation Program for Quantum Science and Technology (Project 2021ZD0302100). 
Z.C.G. is supported by funding from Hong Kong's Research Grants Council (NSFC/RGC Joint Research Scheme No. N-CUHK427/18, GRF No. 14302021 and No. 14301219) and Direct Grant No. 4053462 from The Chinese University of Hong Kong.

\appendix

\section{Fermionic tensor network and $\mathbb{Z}_2$-graded structure}

The study of fermionic tensor networks has become increasingly popular during the last decade.
Several approaches have been proposed, such as GTPS \cite{gu2010grassmann,PhysRevB.88.115139}, the fermionic tensor network with $\mathbb{Z}_2$-graded vector space \cite{PhysRevB.95.075108,Bultinck_2017}, and the fermionic swap gate for the bosonic tensor network \cite{PhysRevB.80.165129,PhysRevA.81.010303,PhysRevB.81.165104,PhysRevA.81.052338}.
These fermionic tensor network approaches are essentially equivalent, and we use the fermionic tensor network with $\mathbb{Z}_2$-graded vector space in this paper.

To describe the fermionic parity symmetry $\mathbb{Z}_2^f$, we introduce the supervector space $V$ with a $\mathbb{Z}_2$-graded structure
\begin{equation}
V = V^0 \oplus V^1, 
\end{equation}
where $|i) \in V^0$ ($V^1$) is an vector with the even (odd) parity denoted by $|i| = 0\,(1)$.
The parity of $|\{i_n\})$ in a rank-$N$ tensor consisting of $V_1 \otimes_{\mathfrak{g}} V_2 \otimes_{\mathfrak{g}} \cdots \otimes_{\mathfrak{g}} V_N$ is determined by $(|i_1| + |i_2| + \cdots + |i_N|)\mod 2$.

To establish the relation between the fermionic parity symmetry and the $\mathbb{Z}_2$-graded vector space, we introduce two rules that supervector spaces need to follow.
The first one is the tensor product isomorphism $\mathcal{F}$:
\begin{equation}
\begin{aligned}
\mathcal{F}: & \quad V_1 \otimes_{\mathfrak{g}} V_2 \rightarrow  V_2 \otimes_{\mathfrak{g}} V_1 : \\ 
& |i_1) \otimes_{\mathfrak{g}} |i_2) \rightarrow  (-1)^{|i_1||i_2|} |i_2) \otimes_{\mathfrak{g}} |i_1),
\end{aligned}
\end{equation}
where even (odd) parity vectors can be viewed as carrying even (odd) fermionic numbers.
The second one is the tensor contraction $\mathcal{C}$, which maps a tensor in $V^* \otimes_{\mathfrak{g}} V$ to $\mathbb{C}$:
\begin{equation}
\mathcal{C}:\quad V^* \otimes_{\mathfrak{g}} V \rightarrow \mathbb{C}: (i'| \otimes_{\mathfrak{g}} |i) = \delta_{i',i}.
\end{equation}
In general, all $\mathbb{Z}_2$-graded tensors are set to satisfy the parity conservation constraint $(|i_1| + |i_2| + \cdots + |i_N|)\mod 2 = 0$.
For instance, the rank-$2$ tensor (matrix) has the block form
\begin{equation}
\begin{aligned}
V_1 \otimes_{\mathfrak{g}} V_2 &= \left( \begin{array}{cc} V_1^0\otimes_{\mathfrak{g}} V_2^0 & V_1^0\otimes_{\mathfrak{g}} V_2^1 \\ V_1^1\otimes_{\mathfrak{g}} V_2^0 & V_1^1\otimes_{\mathfrak{g}} V_2^1 \end{array}\right) \\
&= (V_1^0\otimes_{\mathfrak{g}} V_2^0) \oplus (V_1^1\otimes_{\mathfrak{g}} V_2^1).
\end{aligned}
\end{equation}

One of the most challenging tasks for applying fermionic tensor networks in practice is to perform matrix decompositions.
Taking the $\mathbb{Z}_2^f$-symmetric tensor $\mathcal{T}_{i_1\cdots i_N}$ with even parity as an example, we calculate the SVD decomposition $\mathcal{T} = \mathcal{U} \mathcal{S} \mathcal{V}$ in the following steps.
First, we permute indices and separate them into two groups $I = i_1 \cdots i_n$ and $J = i_{n+1} \cdots i_N$.
We reshape each group to form a large index, i.e. converting the tensor $\mathcal{T}_{i_1 \cdots i_N}$ into a matrix $\mathcal{T}_{I J}$.
The parities $|I|$ and $|J|$ can be rewritten as $|I| = (|i_1| + \cdots + |i_n|) \mod 2$ and $|J| = (|i_{n+1}| + \cdots + |i_N|) \mod 2$.
The matrix contains two blocks $\mathcal{T}_{IJ}= \mathcal{T}^{e}\oplus \mathcal{T}^{o}$, where $|I| = |J| = 0\,(1)$ denotes the even (odd) block $\mathcal{T}^{e}$ ($\mathcal{T}^{o}$).
Next, we make a SVD decompostion respectively for each block, 
\begin{equation}
\begin{aligned}
\mathcal{T}^e_{I_eJ_e} &= \mathcal{U}^e_{I_eK_e} \mathcal{S}^e_{K_eK_e} \mathcal{V}^e_{K_e J_e}, \\
\mathcal{T}^o_{I_oJ_o} &= \mathcal{U}^o_{I_oK_o} \mathcal{S}^o_{K_oK_o} \mathcal{V}^o_{K_o J_o}.
\end{aligned}
\end{equation}
We obtain $\mathcal{U}_{IK} = \mathcal{U}^e \oplus \mathcal{U}^o$, $\mathcal{S}_{KK} = \mathcal{S}^e \oplus \mathcal{S}^o$, and $\mathcal{V}_{KJ} = \mathcal{V}^e \oplus \mathcal{V}^o$.
Finally, we reshape the matrices $\mathcal{U}_{IK}$, $\mathcal{S}_{KK}$, and $\mathcal{V}_{KJ}$ back to the tensor form $\mathcal{U}_{i_1\cdots i_n,k}$, $\mathcal{S}_{kk}$, and $\mathcal{V}_{k,i_{n+1}\cdots i_N}$, and they satisfy $\mathcal{T}_{i_1\cdots i_N} = \mathcal{U}_{i_1\cdots i_n,k} \mathcal{S}_{kk} \mathcal{V}_{k,i_{n+1}\cdots i_N}$.
Other matrix decompositions, such as QR decomposition and polar decomposition, are similar to the process of SVD decomposition.

In the thermodynamic limit, the tensor-network state on a honeycomb lattice is composed of regularly repeated supercells.
Each supercell contains $L_1 \times L_2 \times 2$ distinct rank-$4$ tensors $\mathcal{T}_{[x,y,A]}$ and $\mathcal{T}_{[x,y,B]}$.
Under the fermionic tensor network formalism, we express the tensor-network state as
\begin{equation}
|\Psi\rangle = \mathcal{C}_\nu(\prod_r \mathcal{T}_{[r,A];a_1b_1c_1}^{m_1} \mathcal{T}_{[r,B];a_2b_2c_2}^{m_2} \Lambda_{[r]}^{a_1a_2}\Lambda_{[r]}^{b_1b_2}\Lambda_{[r]}^{c_1c_2}),
\end{equation}
where $r = (x,y)$, and each fermionic tensor is expressed in supervector space as 
\begin{align}
\mathcal{T}_{[r,A];a_1b_1c_1}^{m_1} &= T_{[r,A];a_1b_1c_1}^{m_1} |m_1\rangle (a_1|(b_1|(c_1|  \notag\\
& \in \mathcal{H}_{[r,A]} \otimes_{\mathfrak{g}} V_{[r,A];a_1}^*\otimes_{\mathfrak{g}} V_{[r,A];b_1}^*\otimes_{\mathfrak{g}} V_{[r,A];c_1}^* , \notag\\
\mathcal{T}_{[r,B];a_2b_2c_2}^{m_2} &= T_{[r,B];a_2b_2c_2}^{m_2} |m_2\rangle (a_2|(b_2|(c_2|  \notag\\
& \in \mathcal{H}_{[r,B]} \otimes_{\mathfrak{g}} V_{[r,B];a_2}^*\otimes_{\mathfrak{g}} V_{[r,B];b_2}^*\otimes_{\mathfrak{g}} V_{[r,B];c_2}^* ,
\end{align}
\begin{align}
\Lambda_{[r]}^{a_1a_2} =&\, \lambda_{a_1 a_2} |a_1)|a_2) \in V_{[x,y,A];a_1} \otimes_{\mathfrak{g}} V_{[x,y-1,B];a_2}, \notag\\
\Lambda_{[r]}^{b_1b_2} =&\, \lambda_{b_1 b_2} |b_1)|b_2) \in V_{[x,y,A];b_1} \otimes_{\mathfrak{g}} V_{[x-1,y,B];b_2}, \\
\Lambda_{[r]}^{c_1c_2} =&\, \lambda_{c_1 c_2} |c_1)|c_2) \in V_{[x,y,A];c_1} \otimes_{\mathfrak{g}} V_{[x,y,B];c_2}. \notag
\end{align}
Here, the supervector spaces $\mathcal{H}_{[r,A]}$ and $\mathcal{H}_{[r,B]}$ indicate physical spaces.
The supervector space $V_{[r,A(B)];\alpha=\{a,b,c\}}$ represents the virtual index on the tensor $\Lambda_{[r]}$, while the corresponding supervector space $V_{[r,A(B)];\alpha}^*$ represents the virtual index on the tensor $\mathcal{T}_{[r,A(B)]}$.
The contraction map $\mathcal{C}$ yields $(\alpha' |\alpha ) = \delta_{\alpha\alpha'}$, where $|\alpha) \in V_{[r,A(B)];\alpha}$ and $(\alpha'| \in V_{[r,A(B)];\alpha}^*$.
The elements $T_{[r,A];a_1b_1c_1}^{m_1}$ and $\lambda_{a_1 a_2}$ are coefficients of fermionic tensors.
The rank-$2$ tensor $\Lambda_{[r]}$ denotes the Schmidt weight between two sub-lattice sites.
$\mathcal{T}_{[r,A];a_1b_1c_1}^{m_1}$ and $\mathcal{T}_{[r,B];a_2b_2c_2}^{m_2}$ are rank-$4$ tensors with three virtual indices and one physical index $|m\rangle$, which has three possible states: one hole state $\left|0\right\rangle$ with even parity and two electronic states $\left|\uparrow\right\rangle$ and $\left|\downarrow\right\rangle$ with odd parity.
All tensors satisfy the even parity conservation condition, e.g., $(|m_{1(2)}| + |a_{1(2)}| + |b_{1(2)}| + |c_{1(2)}|) \mod 2 = 0$ for $\mathcal{T}_{[r,A];a_1b_1c_1}^{m_1}$ and $\mathcal{T}_{[r,B];a_2b_2c_2}^{m_2}$.

\section{More Details on Uniform States}
\label{sec:MRandD}

\begin{figure*}[tbp]
    \centering
    \includegraphics[width=2.0\columnwidth]{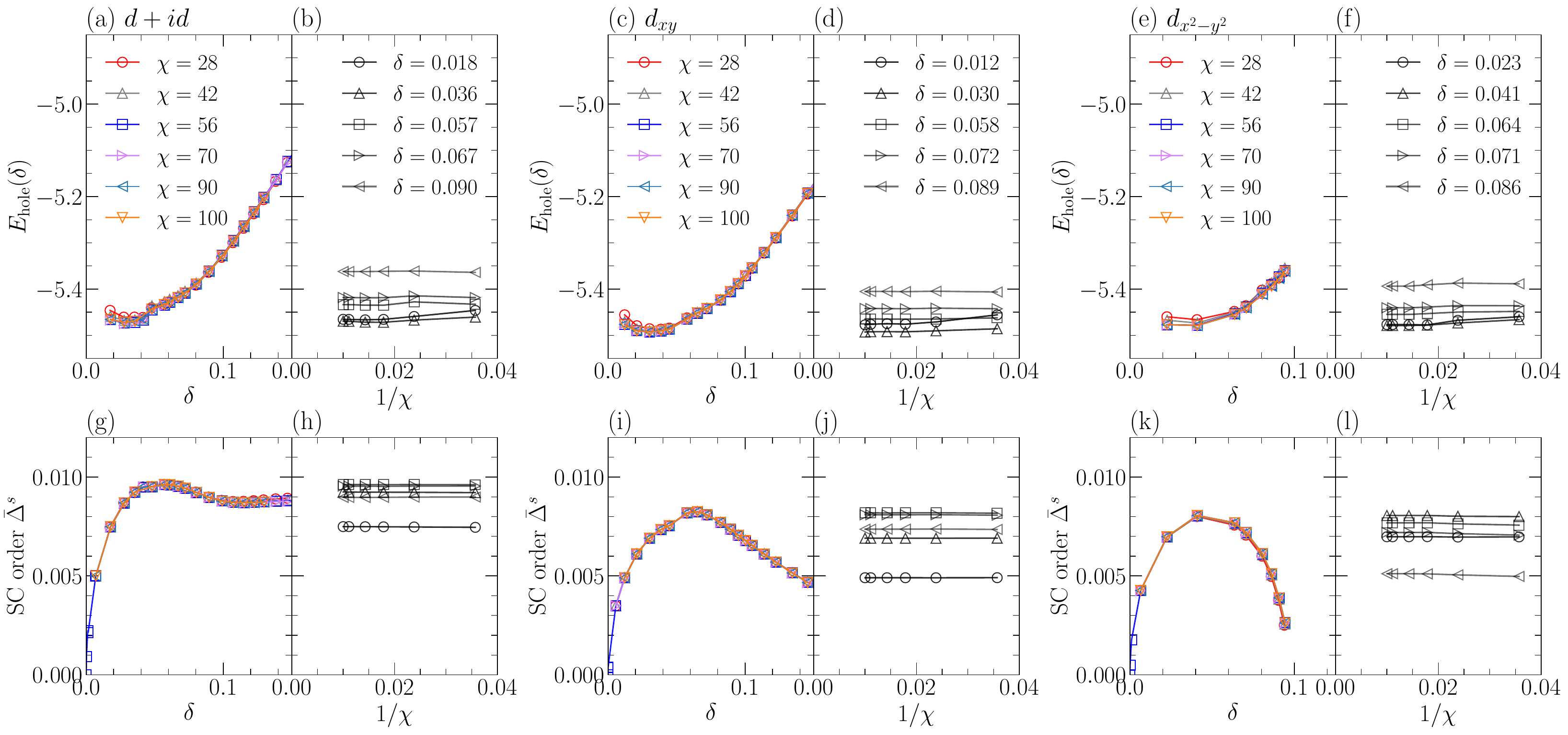} 
    \caption{
    Results for three uniform $d$-wave states with $D=14$ and various environmental bond dimensions $\chi$.
    (a), (c), (e) The energy per hole $E_{\mathrm{hole}}(\delta)$.
    (b), (d), (f) The $1/\chi$ scaling data of $E_{\mathrm{hole}}(\delta)$.
    The energies are convergent with $\chi \gtrsim 4D$ at $\delta > 0.01$.
    (g, i, k) The SC order $\bar{\Delta}^s$.
    (h, j, l) The $1/\chi$ scaling data of $\bar{\Delta}^s$.
    }
    \label{fig:7}
\end{figure*}

\begin{figure}[tbp]
    \centering
    \includegraphics[width=1.0\columnwidth]{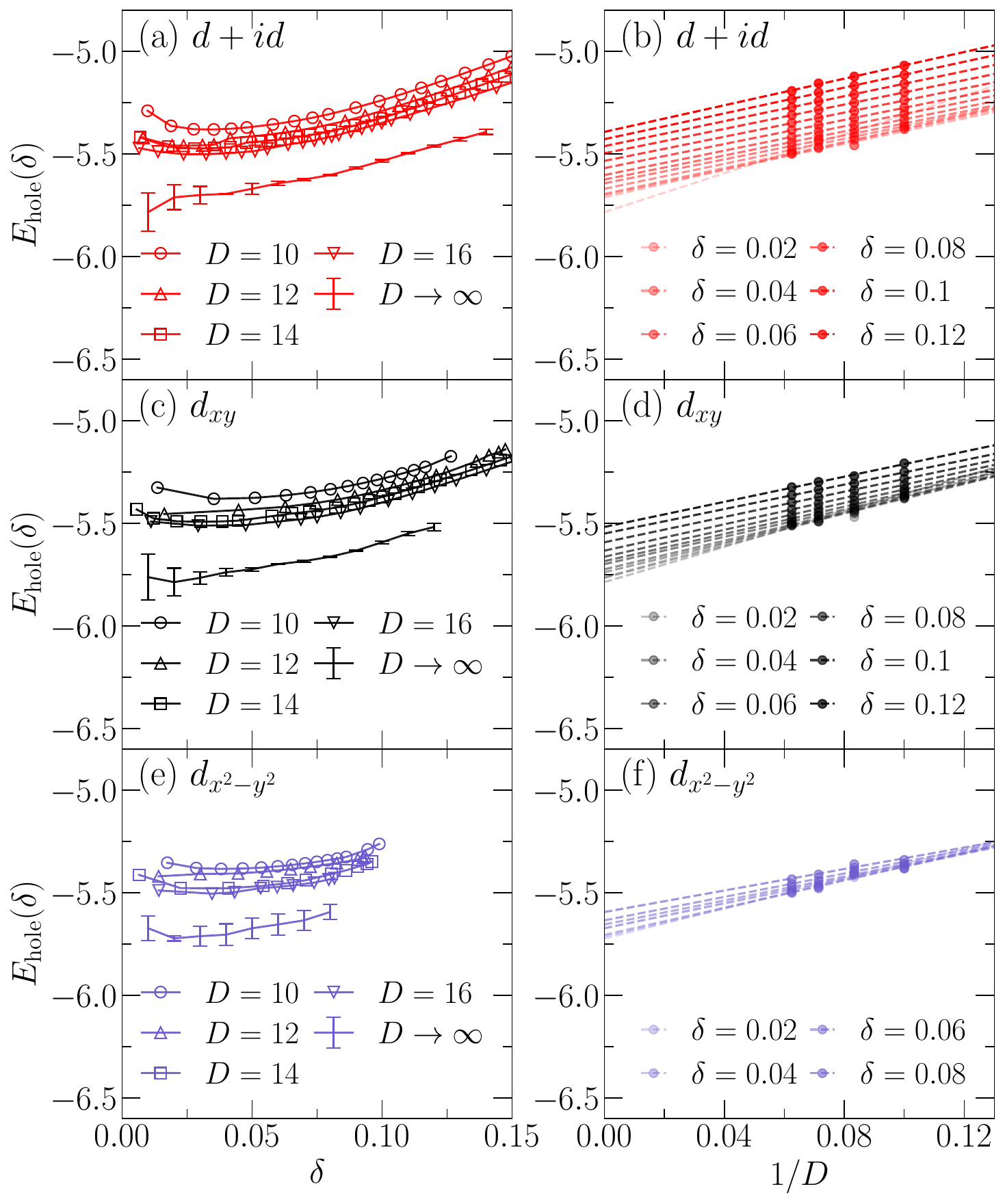} 
    \caption{
    The extrapolation of the energy per hole for three uniform states.
    (a), (c), (e) Energy per hole $E_{{\rm hole}}(\delta)$.
    (b), (d), (f) Details of linear extrapolation with $1/D$.
    We find that the energy approximately satisfies $E(D) = e_0 + \alpha D^{-1}$ for various hole dopings.
    }
    \label{fig:8}
\end{figure}

\begin{figure}[tbp]
    \centering
    \includegraphics[width=1.0\columnwidth]{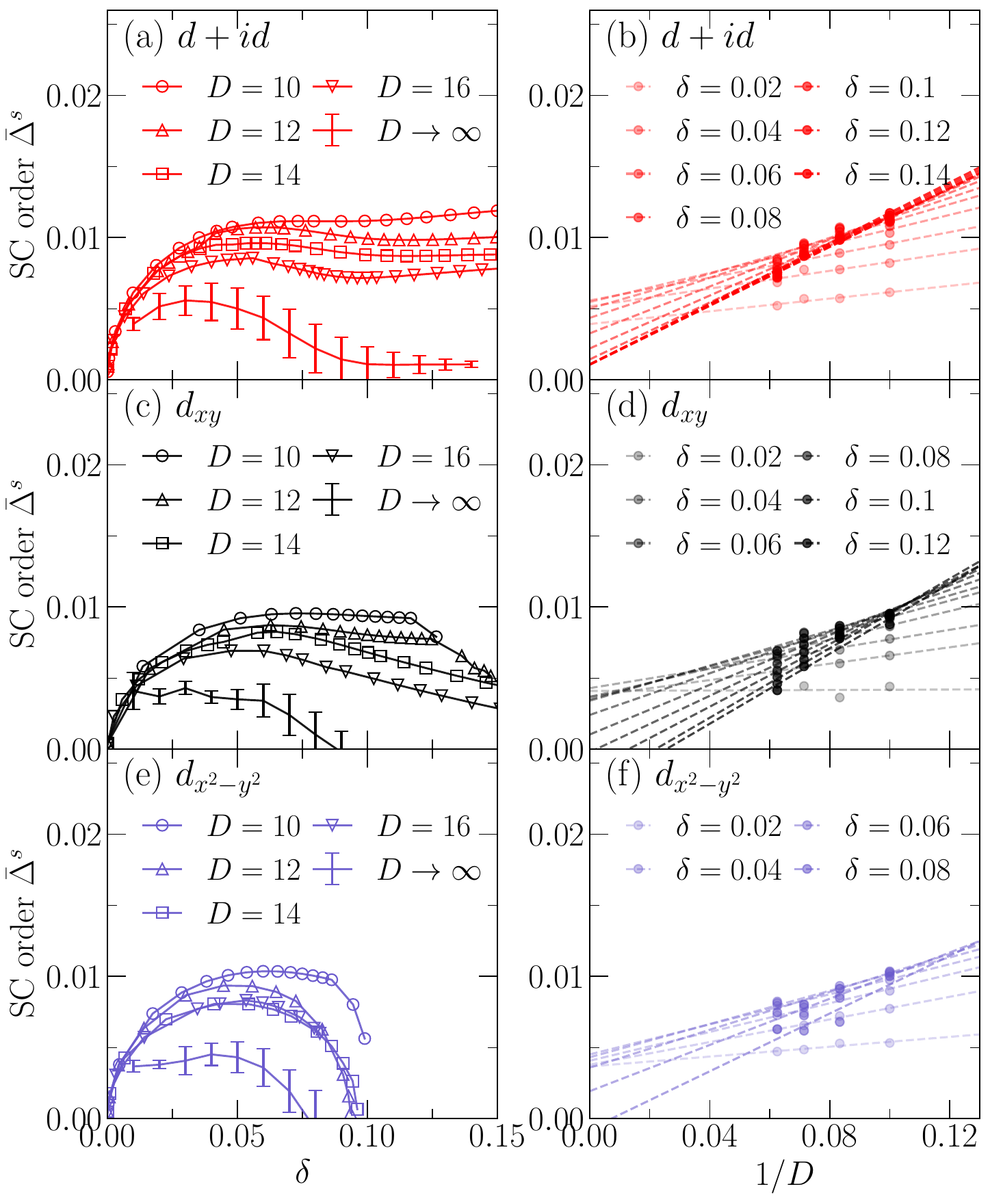} 
    \caption{
    The extrapolation of the SC order for three uniform states.
    (a), (c), (e) The SC order.
    (b), (d), (f) Details of the linear extrapolation with $1/D$.
    }
    \label{fig:9}
\end{figure} 

\begin{figure}[tbp]
    \centering
    \includegraphics[width=1.0\columnwidth]{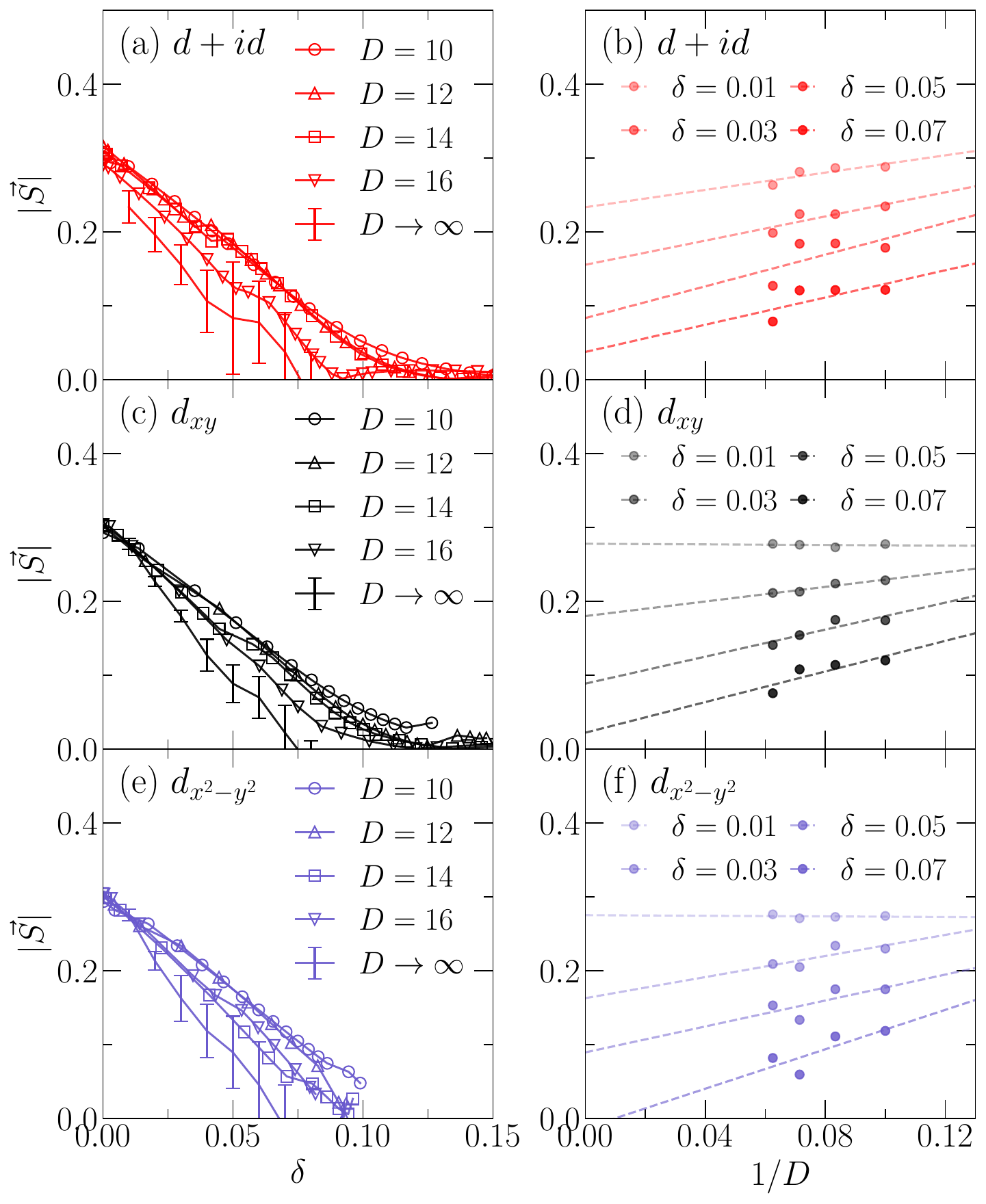} 
    \caption{
    The extrapolation of the staggered magnetization for three uniform states.
    (a), (c), (e) The staggered magnetization.
    (b), (d), (f) Details of the linear extrapolation with $1/D$.
    }
    \label{fig:10}
\end{figure} 

\begin{figure}[tbp]
    \centering
    \includegraphics[width=0.9\columnwidth]{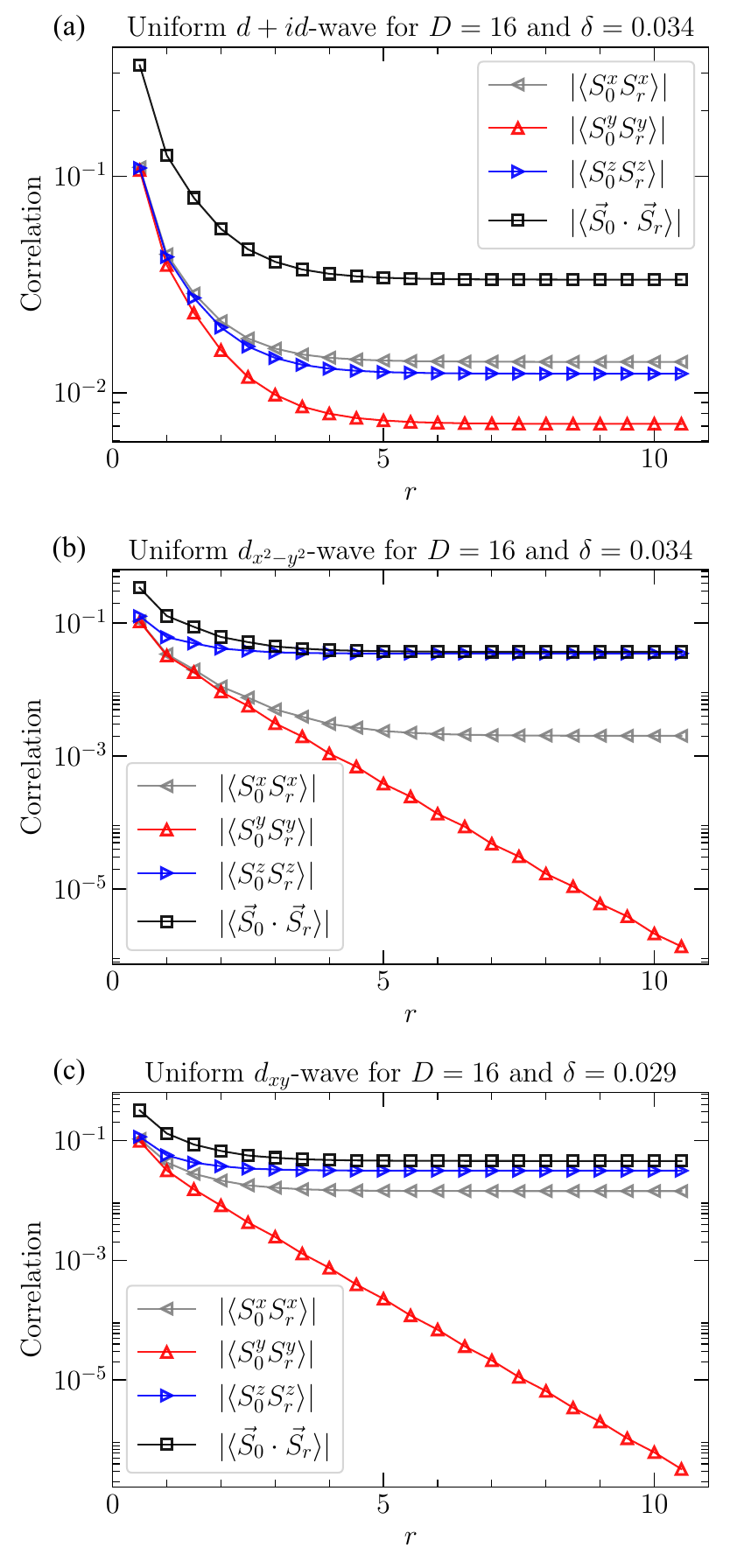} 
    \caption{
        Spin-spin correlation functions of three uniform states for $D=16$ and $\delta \approx 0.03$.
    }
    \label{fig:11}
\end{figure}

\begin{figure}[tbp]
\centering
\includegraphics[width=1.0\columnwidth]{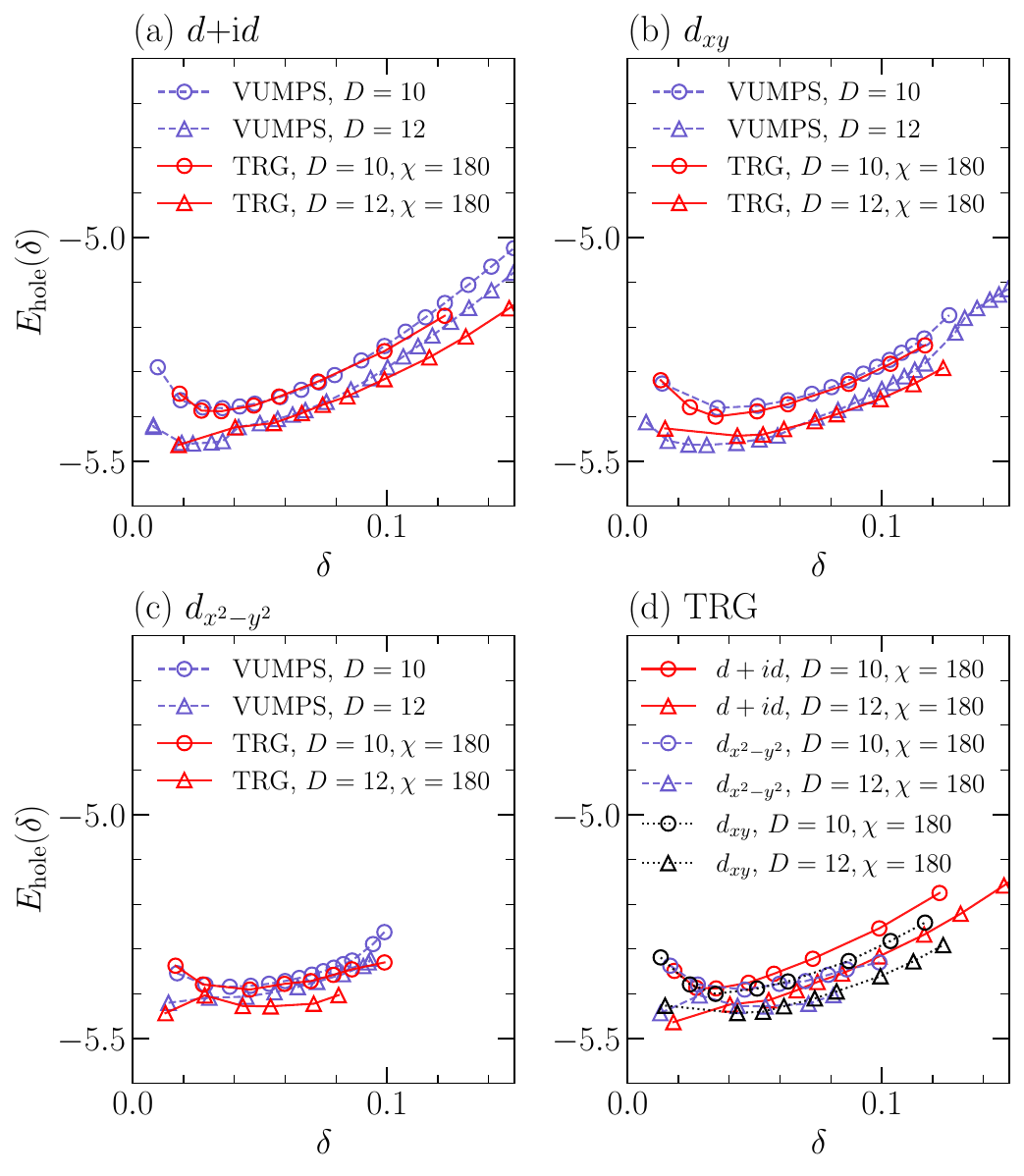} 
\caption{
Energy comparison between two methods, VUMPS and TRG, for three uniform states.
(a) $d+id$-wave uniform states, 
(b) $d_{xy}$-wave uniform states, 
(c) $d_{x^2-y^2}$-wave uniform states. 
Two methods yield almost identical results with $D=10$ and $12$, indicating the reliability of calculations.
(d) Energies for three uniform states in the TRG calculation. 
Three uniform states are nearly degenerate.
}
\label{fig:12}
\end{figure}

\begin{figure}[tbp]
\centering
\includegraphics[width=1.0\columnwidth]{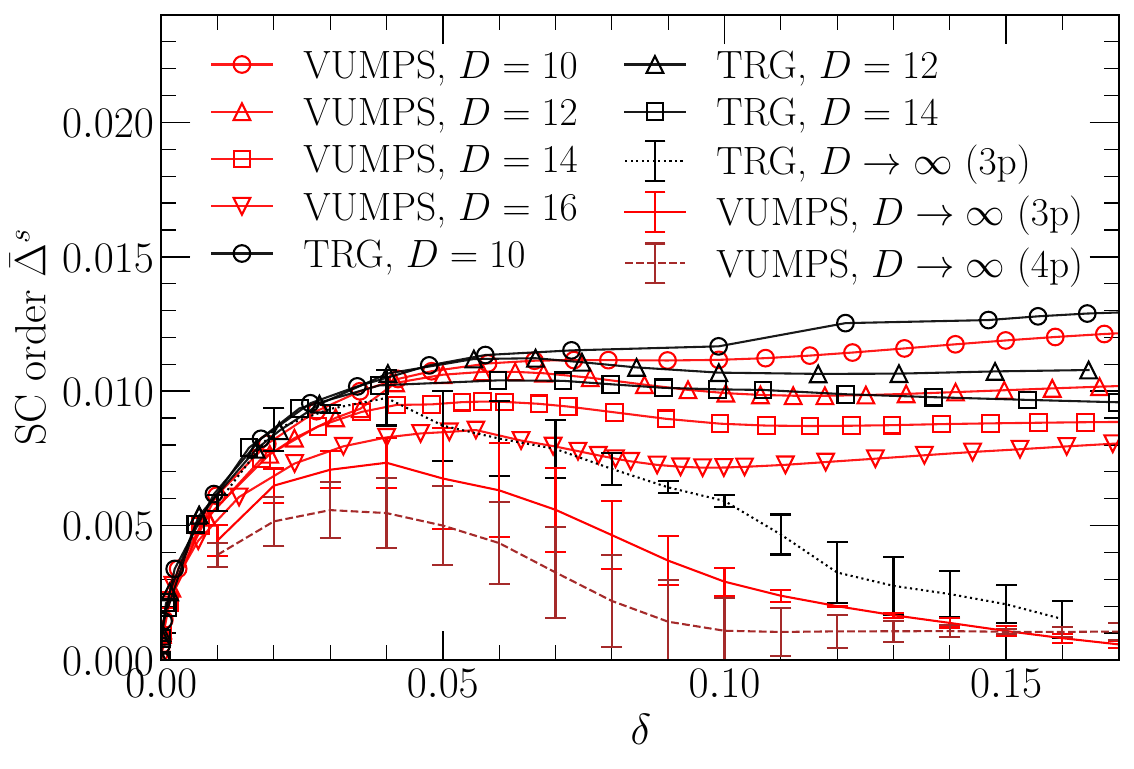} 
\caption{
Comparison of the SC order for $d+id$-wave uniform states between two methods, VUMPS and TRG. 
For $D\rightarrow \infty$, the line labeled by ``3p'' is extrapolated  using three data points with $D=10,12,14$, and the line labeled ``4p'' is extrapolated using four data points with $D=10,12,14,16$. 
Generally, the SC order from TRG calculation is larger than the one from VUMPS calculation for each $D$ and the extrapolation.
}
\label{fig:13}
\end{figure}

In this process, we employ the VUMPS algorithm \cite{10.21468/SciPostPhysLectNotes.7, PhysRevB.97.045145, PhysRevB.98.235148} to contract the two-dimensional tensor network to calculate physical quantities.
To assure the convergence of physical quantities with environmental bond dimension $\chi$, we compute the energy per hole and the SC order of uniform $d$-wave states with $D=14$ and various $\chi$ in Fig. \ref{fig:7}.
When $\chi \gtrsim 4 D$, the energy per hole appears to be convergent, and so does the SC order.
    
Using $D=10,12,14,16$ results, we perform a linear extrapolation in $1/D$ for various physical quantities.
Figure \ref{fig:8} depicts the extrapolation of the energy per hole for three uniform states and details about the linear extrapolation. 
In Figs. \ref{fig:8}(b), \ref{fig:8}(d) and \ref{fig:8}(f), the energies per hole for three uniform states satisfy $E(D) = e_0 + \alpha D^{-1}$.
In the extrapolation, we also discover that three uniform states are nearly degenerate at small hole doping $\delta < 0.06$.

When the energy per site exhibits a convex relationship with respect to hole doping, it indicates that the system maintains stability.
This suggests that as hole doping increases, the energy per hole will correspondingly escalate in a linear manner. 
If the energy per hole reaches its minimum at a specific doping level, denoted as $\delta_c$, phase separation transpires for hole doping $\delta$ ranging between $0$ and $\delta_c$ \cite{PhysRevLett.113.046402,PhysRevLett.64.475}.
As shown in Figs. \ref{fig:8}(a) and \ref{fig:8}(c), we can see the minima of energy per hole is approximately close to half filling for uniform $d+id$-wave and $d_{xy}$-wave states in the extrapolation with $1/D$, indicating no phase separation for these two states.
At $\delta > 0.05$, we observe the possible $\delta_c = 0.06$ for W8 stripe state in Fig. \ref{fig:4}, which suggests phase separation for $\delta < 0.06$. 
However, more detailed studies for stripe states at larger $D$ are required to handle the issue.
    
For the SC order, we also make a linear extrapolation in $1/D$ in Fig. \ref{fig:9}.
The SC order of the $d_{x^2-y^2}$-wave state falls rapidly at $\delta \approx 0.1$ for each $D$, and the extrapolation has a small hole doping survival region.
The SC order of the $d_{xy}$-wave state exists at $\delta > 0.15$ for each $D$, but there is no SC order in the extrapolation at $\delta > 0.09$.
Unlike the other two states, the SC order of the $d+id$-wave can exist at large hole doping, even with $\delta > 0.15$ in extrapolation.
At $\delta > 0.1$, neither the $d_{xy}$-wave nor the $d_{x^2-y^2}$-wave state has an SC order, but the $d+id$-wave state has an SC order that survives over a wide range of hole doping.

As shown in Fig. \ref{fig:10}, we linearly extrapolate the staggered magnetizations with $1/D$.
In the extrapolation, we find that the magnetizations of the three uniform states still decrease (approximately) linearly with increasing hole doping $\delta$ and vanish at $\delta \gtrsim 0.07$.

From the correlation function perspective, we perform further calculations to analyze the spin order, similar to the analysis of spin symmetry in the square $t-J$ model using iPEPS \cite{PhysRevB.103.075127}.
As illustrated in Fig. \ref{fig:11}, we present the spin-spin correlation function $\langle \vec{S}_0\cdot\vec{S}_r\rangle$ for three uniform states and its corresponding components $\langle S^x_0 S^x_r\rangle$, $\langle S^y_0 S^y_r\rangle$, and $\langle S^z_0 S^z_r\rangle$. 
In the thermodynamic limit, as $r$ increases, $\langle \hat{A}_0 \hat{A}_r\rangle$ approaches $\langle \hat{A}_0\rangle \langle \hat{A}_r\rangle$, and a non-zero $\langle \hat{A}_i \rangle$ gives rise to long-range order.
For the three uniform states, it is evident that spin-spin correlations persist over long distances in Fig. \ref{fig:11}. 
As depicted in Fig. \ref{fig:10}, the magnetizations of the three uniform states are present at $\delta < 0.07$ in the extrapolation of magnetization with $1/D$, suggesting that the coexisting AFM states exhibit long-range order. 
Furthermore, we observe that for the uniform $d_{xy}$-wave and $d_{x^2-y^2}$-wave states, $\langle \hat{S}^y_i\rangle = \langle 1j(c_{i\downarrow}^\dagger c_{i\uparrow} - c_{i\uparrow}^\dagger c_{i\downarrow})/2\rangle= 0$, owing to the choice of real wave functions for these two states. 
This observation indicates that $\langle\hat{S}^y_0 \hat{S}^y_r\rangle$ approaches $\langle\hat{S}^y_0\rangle \langle \hat{S}^y_r\rangle = 0$ as $r$ increases.
In contrast, for $d+id$-wave states with complex wave functions, $\langle\hat{S}^y_0 \hat{S}^y_r\rangle \neq 0$.

In the coexisting AFM state, the spins align in an antiferromagnetic pattern that repeats precisely every two sites.
This alignment results from the tensor-network state Ansatz, which consists of only two tensors in the thermodynamic limit. 
Consequently, our numerical method yields a commensurate AFM phase for the coexisting AFM state.

In addition, we use the TRG approach to contract the two-dimensional scalar product $\langle \Psi |\Psi \rangle$.
The energies obtained by TRG for three uniform states with $D=10,12$ are shown in Figs. \ref{fig:12}(a)-\ref{fig:12}(c), and the truncation dimension for TRG is set to $\chi = 180$.
The energies of three uniform states in TRG calculation are nearly identical to those in VUMPS calculation, indicating the reliability of the two algorithms.
In Fig. \ref{fig:12}(d), the energies for the three states are approximately equal at $\delta < 0.05$ according to the TRG approach.
    
Using TRG and VUMPS, we compute the SC order for the uniform state with $d+id$-wave pairing symmetry, as shown in Fig. \ref{fig:13}.
The SC order in the TRG calculation is slightly larger than the SC order in the VUMPS calculation for each $D$ and the extrapolation.
We can see that extrapolation in TRG calculation results in the existence of SC order at $\delta > 0.15$.

\section{More Details on $d_{x^2-y^2}$-wave Stripe States}
\label{sec:dx2SS}

\begin{figure}[tbp]
    \centering
    \includegraphics[width=1.0\columnwidth]{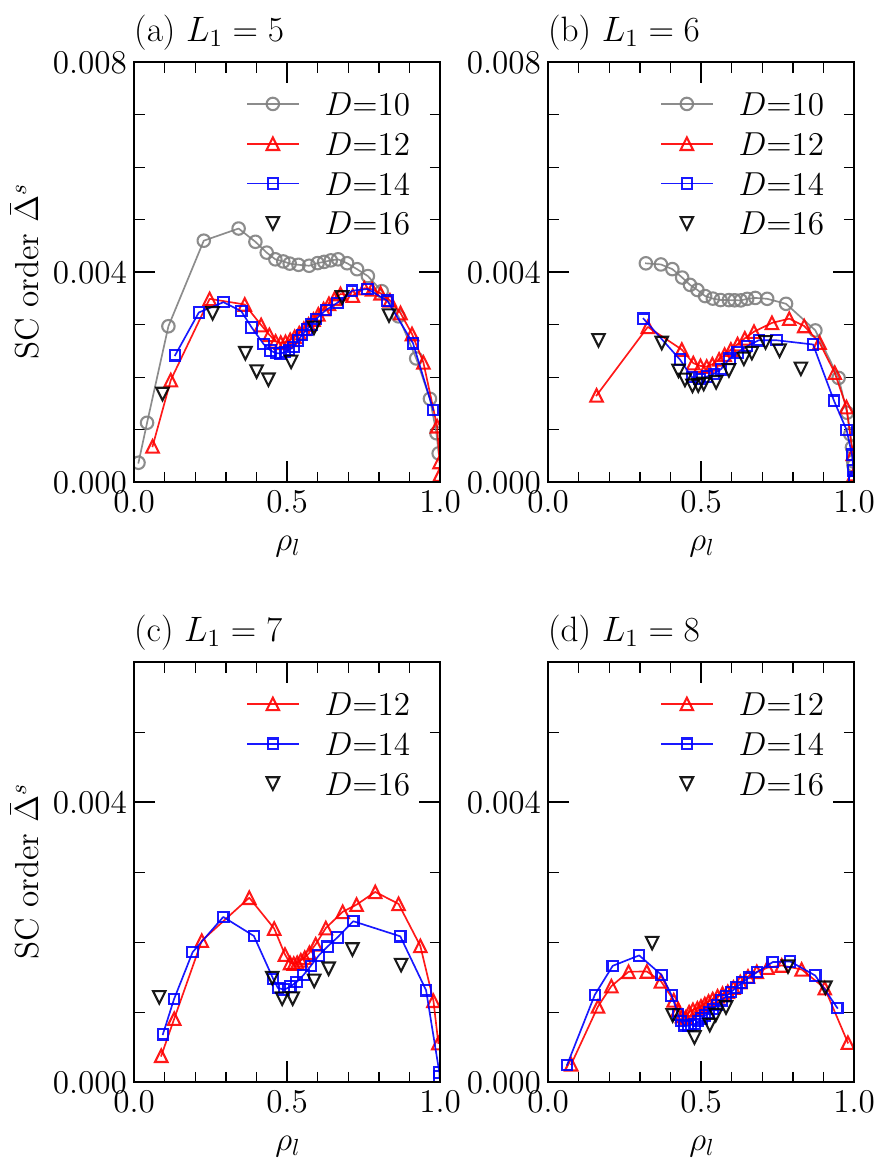} 
    \caption{
    The SC order for $d_{x^2-y^2}$-wave stripe states with different stripe periods by increasing $D$:
    (a) $L_1 = 5$, (b) $L_1=6$, (c) $L_1=7$ and (d) $L_1 = 8$.
    These stripe states display AFM order, similar to uniform states, but have higher energies than uniform states at $\rho_l < 0.5$.
    Two peaks in the SC order with changing $\rho_l$ characterize the transition from uniform-like states to stripe states with a $\pi$-phase shift of the staggered magnetization.
    }
    \label{fig:14}
\end{figure} 

\begin{figure}[tbp]
    \centering
    \includegraphics[width=1.0\columnwidth]{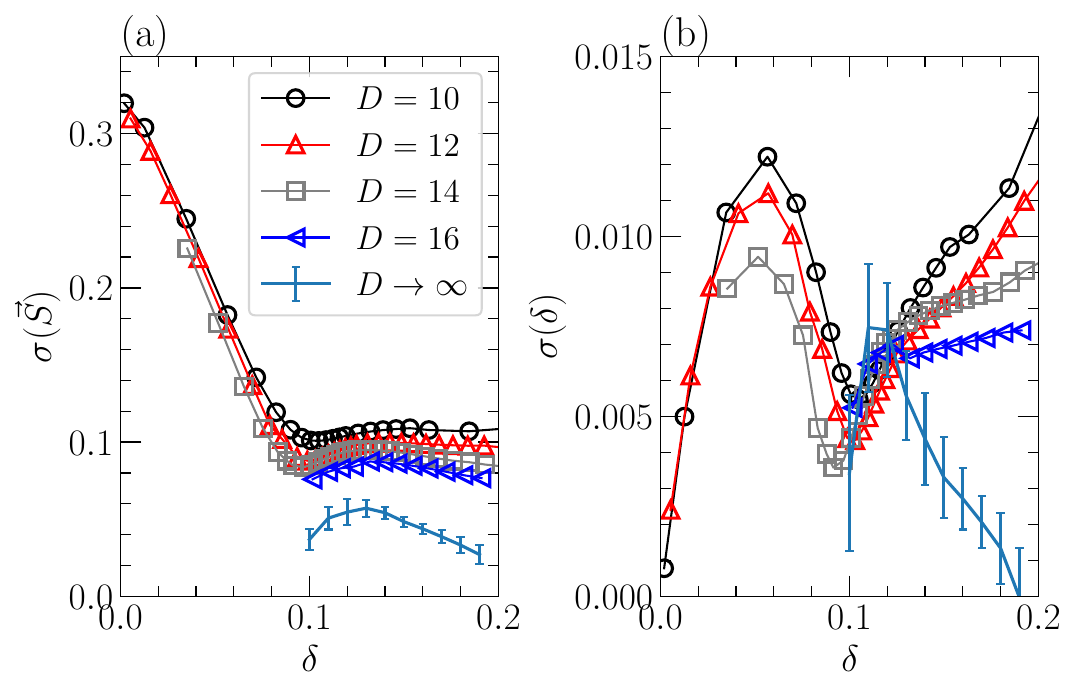} 
    \caption{
        Results for W4 stripe states.
        (a) The variance of the magnetization $\mathrm{var}(\vec{S})$.
        (b) The variance of the local hole density $\mathrm{var}(\delta)$.
    }
    \label{fig:15}
\end{figure}

To demonstrate the existence of the SC order for $d_{x^2-y^2}$-wave stripe states, we show the SC order as $D$ increases in Fig. \ref{fig:14}.
We observe that the SC order with $D=12$ is rather close to the SC order with $D=14$ for each class of states with different stripe periods, and results with the larger $D=16$ also indicate the existence of the SC order for $d_{x^2-y^2}$-wave stripe states.
Furthermore, unlike states with a $\pi$-phase shift of the staggered magnetization at $\rho_l > 0.5$, states at $\rho_l < 0.5$ show an AFM order, which is similar to uniform states but with higher energies.
Two peaks in the SC order with a change of $\rho_l$ characterize the transition from uniform-like states to stripe states.

To investigate the stripe order, we examine various states across a range of hole doping levels and unit cell sizes.
The phase diagram in Fig. \ref{fig:1} displays a range of $\delta$ values from 0 to 0.17. 
Our findings indicate that the lowest energy state within the interval $\delta\in[0.12,0.17]$ corresponds to the $d_{x^2-y^2}$-wave W4 stripe state, as illustrated in Fig. \ref{fig:1}(c) in the main text. 
Moreover, we observe that when $\delta$ exceeds 0.17, the energies of states with periods 2 and 3 approach or become lower than that of the W4 stripe state. 
This observation suggests a gradual transition toward uniform states for $\delta > 0.17$.
As our primary focus is on the stripe order, we have limited the phase diagram interval to $\delta$ values between 0 and 0.17. 
We notice that magnetization decreases with increasing hole doping levels, while charge order emerges at $\delta$ values greater than 0.05.
To exclude the finite $D$ effect of the stripe order for $\delta > 0.1$ and make an extrapolation of the stripe order, we use the following expressions to characterize it
\begin{equation}
\sigma(\vec{S}) = \sqrt{\frac{1}{2L_1}\sum_{i}\langle (\vec{S}_i - \bar{S})^2 \rangle},
\end{equation}
\begin{equation}
\sigma(\delta) = \sqrt{\frac{1}{2L_1}\sum_{i}\langle (\delta_i - \bar{\delta})^2 \rangle}.
\end{equation}
Here, $\bar{S} = \frac{1}{2L_1}\sum_{i} \langle \vec{S}_i \rangle$, $\bar{\delta} = \frac{1}{2L_1}\sum_i \langle \delta_i \rangle$, $i$ denotes the site location, and $L_1 = 4$. 
Figure \ref{fig:15} illustrates the standard deviations of $\vec{S}$ and $\delta$ as a function of $\delta$ for states on $4 \times 1$ unit cell. 
We observed a valley of $\sigma(\vec{S})$ at $\delta_c \approx 0.1$, indicating a transition from a uniform-like AFM state to the W4 stripe state on $4 \times 1$ unit cell. 
In the extrapolation of $\sigma(\vec{S})$ with $1/D$, SDW persists at $\delta \in [0.12,0.17]$ in (a). 
The CDW, identified by $\sigma(\delta)$, also exists at $\delta < 0.19$ in (b), while $\sigma(\delta)$ of a uniform state equals zero. 
These findings suggest that the stripe order does not vanish in the infinite $D$ limit for the W4 stripe state at $\delta \in[0.12,0.17]$.

\section{The $d_{xy}$-wave Stripe States}
\label{sec:dxySS}

\begin{figure}[tbp]
\centering
\includegraphics[width=1.0\columnwidth]{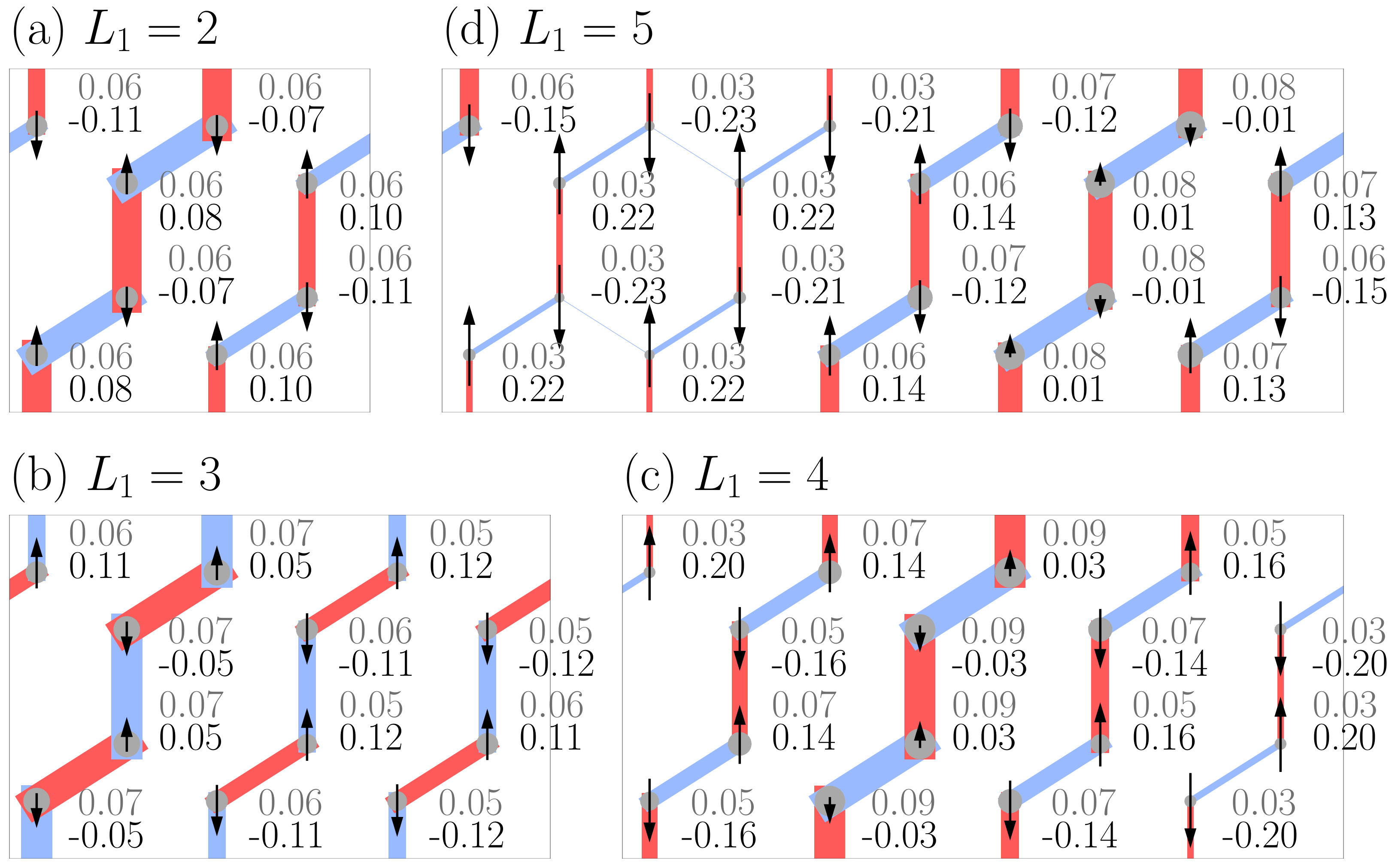} 
\caption{
Patterns of $d_{xy}$-wave stripe states with $D=14$ and $L_1 = 2,3,4,5$ at $\delta \approx 0.06$.
For these stripe states, SC orders along the $a$ and $c$ directions are opposite in sign, and SC orders along the $b$ direction are approximately equal to zero.
Similar to the uniform $d_{xy}$-wave state, $d_{xy}$-wave stripe states are mainly found at low doping with AFM order.
}
\label{fig:16}
\end{figure}

\begin{figure}[tbp]
\centering
\includegraphics[width=1.0\columnwidth]{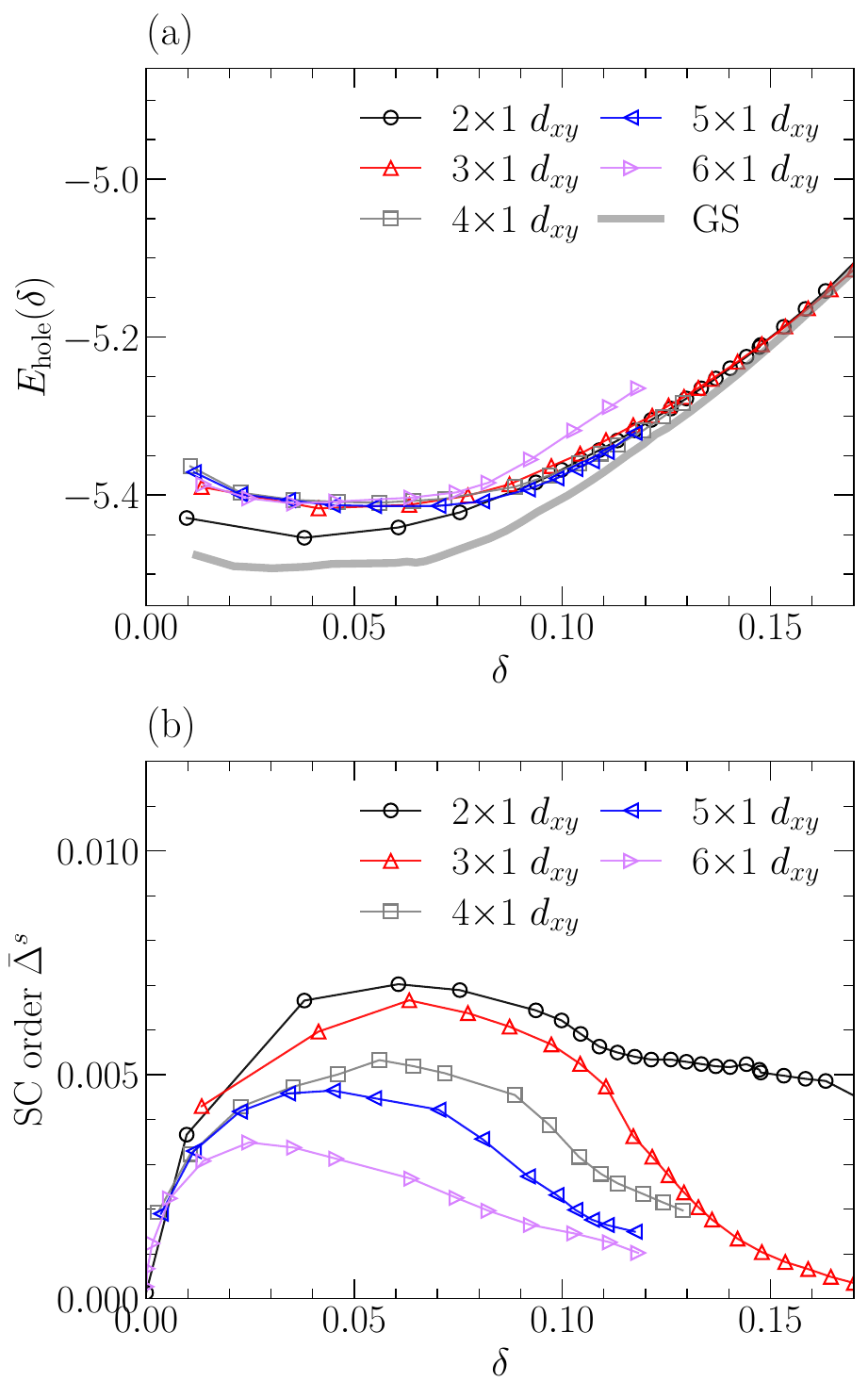} 
\caption{
Results for $d_{xy}$ stripe states with $D = 14$.
(a) The energy per hole $E_{{\rm hole}}(\delta)$.
GS stands for the ground states, as shown in the main text. 
At $\delta < 0.17$, all $d_{xy}$ stripe states have higher energy than the GS.
(b) The SC order $\bar{\Delta}^s$ as a function of hole doping.
The SC order for $d_{xy}$ stripe states also decreases as the stripe period increases.
}
\label{fig:17}
\end{figure} 

\begin{figure}[tbp]
\centering
\includegraphics[width=1.0\columnwidth]{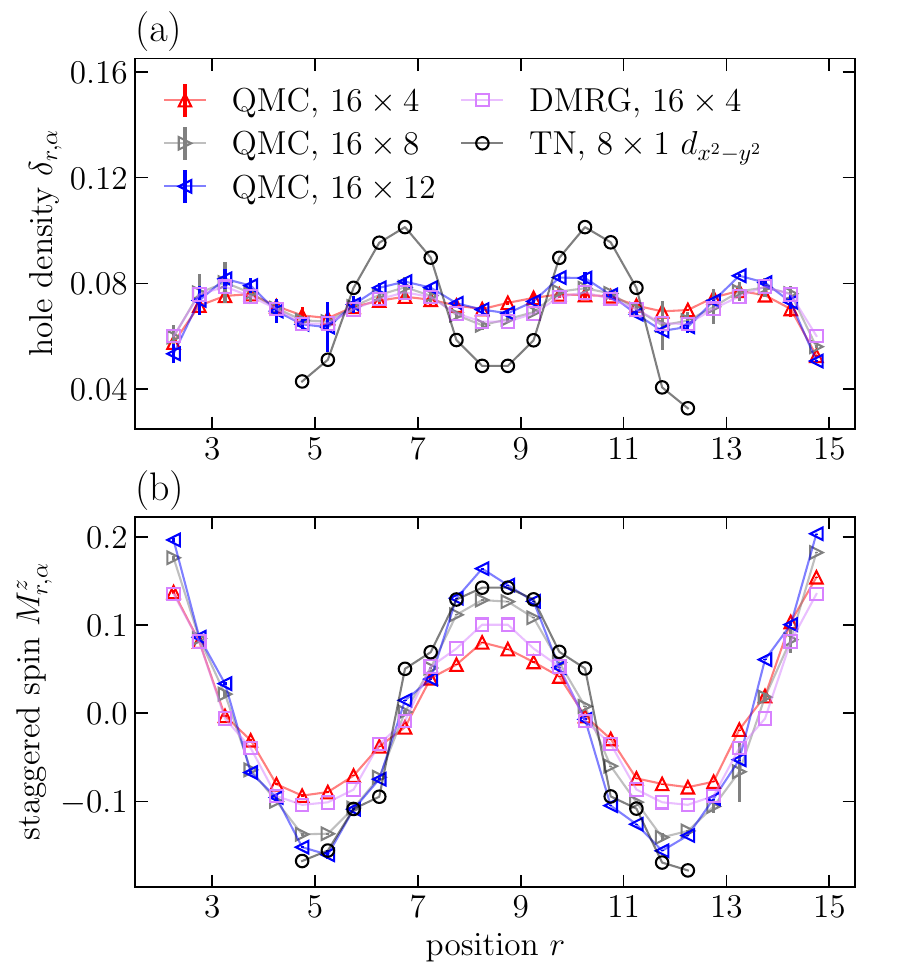} 
\caption{
Results of various approaches for (a) the local hole density $\delta_{r,\alpha}$ and (b) the staggered spin density $M^z_{r,\alpha}$, including Hubbard model with $U/t = 8.0$ and $\delta = 1/16$ on a width-$4$ cylinder using DMRG and width-$4$, $8$, and $12$ cylinders using QMC from Ref. \cite{PhysRevB.105.035111}.
The TN result in our calculation is for the $t$-$J$ model with $J/t = 3.0$ and $\delta \approx 0.07$ on the $L_1 = 8$ supercell, and the state displays $d_{x^2-y^2}$-wave pairing symmetry.
To avoid the boundary effect, we compare the TN result with the QMC/DMRG result in the bulk of the cylinder.
}
\label{fig:18}
\end{figure}

In addition to the $d_{x^2-y^2}$-wave stripe states described in the main text, we also discover stripe states with $d_{xy}$-wave superconducting pairing symmetry. 
We show patterns of $d_{xy}$-wave stripe states on various supercells in Fig. \ref{fig:16}.
The $d_{xy}$-wave stripe state can be described as a state with positive SC orders along the $a$ direction, negative SC orders along the $c$ direction, and zero SC orders along the $b$ direction.
The SC order along the $a$ direction has the same amplitude as the SC order along the $c$ direction from the same site.
Furthermore, for $\delta \gtrsim 0.1$ and $L_1 \geq 4$, the $d_{xy}$-wave pairing symmetry is fragile.
In other words, stripe states with $d_{xy}$-wave pairing symmetry are uncommon at large hole doping.

As shown in Fig. \ref{fig:17}, we calculate the energy per hole and the SC order for the $d_{xy}$-wave stripe states with different stripe periods.
All $d_{xy}$-wave stripe states have higher energy than ground states from uniform states and $d_{x^2-y^2}$-wave stripe states in Fig. \ref{fig:17}(a). 
These stripe states are nearly-degenerate at large hole doping $\delta > 0.14$.
In Fig. \ref{fig:17}(b), we can see that the SC order for $d_{xy}$-wave stripe states mainly occurs at small hole doping, whereas the superconductivity in $d_{x^2-y^2}$-wave stripe states occurs at around $\rho_l \in [0.5,1.0]$.
The pairing amplitude of $d_{xy}$-wave stripe states also decreases as the stripe period increases.
    
A recent numerical calculation \cite{PhysRevB.105.035111} suggests a local $d_{xy}$-wave stripe state with a $\pi$-phase shift of the staggered magnetization at $\delta = 1/16$ on a width-$4$ cylinder using DMRG and width-$4$, $8$ and $12$ cylinders using AFQMC.
We also find similar stripe states that have a $\pi$-phase shift of the staggered magnetization but have $d_{x^2-y^2}$-wave pairing symmetry in the thermodynamic limit.
As shown in Fig. \ref{fig:18}, we compare our result for a W8 $d_{x^2-y^2}$-wave stripe state with the DMRG result and the QMC result from Ref. \cite{PhysRevB.105.035111}.
In Fig. \ref{fig:18}(a), the hole densities for various approaches have a similar trend, but the amplitude of hole modulation for the TN result is larger than that for the DMRG and QMC results because of the stronger interaction or the absence of double-occupancy states.
Staggered spin densities for various approaches are approximately identical in Fig. \ref{fig:18}(b).
However, $d_{xy}$-wave stripe states in our calculation have slightly higher energy than $d_{x^2-y^2}$-wave stripe states.

\begin{figure*}
\centering
\includegraphics[width=2.0\columnwidth]{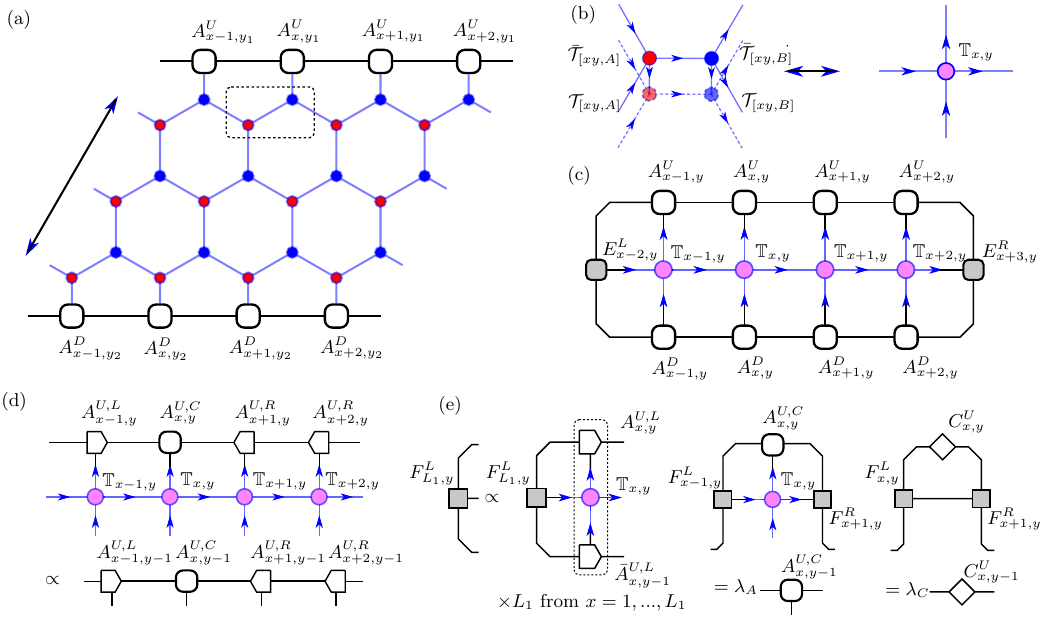} 
\caption{
The boundary MPS method for contracting the scalar product $\langle \Psi|\Psi\rangle$ on the honeycomb lattice.
(a) Contraction in the vertical direction with the boundary MPS $\{ A^U_{x,y}\}$ and $\{ A^D_{x,y}\}$.
(b) The unit component of the scalar product in the form of MPO is obtained by regarding bipartite sites on the honeycomb lattice as one site $\mathbb{T}_{x,y} = \mathcal{C}_v\left(\mathcal{T}_{xy,A}\mathcal{T}_{xy,B}\bar{\mathcal{T}}_{xy,A}\bar{\mathcal{T}}_{xy,B}\right)$.
(c) The left and right fixed points $E^L_{x,y}$ and $E^R_{x,y}$.
(d) The fixed point problem to be solved by VUMPS algorithm.
(e) Key steps for optimizing the overlap.
}
\label{fig:19}
\end{figure*}

\section{Simple Update}
\label{sec:SU}

We choose simple update (SU) to obtain ground states, which is a method to update tensors at the cheapest cost.
The computational cost can be further reduced by applying QR decompositions to tensors before acting on each evolution operator.
For better convergence, we adopt the strategy to start with a large time step and to reduce it as the iteration goes, i.e., $\tau$ gradually decreases from $\tau_{\mathrm{start}} = 0.02$ to $\tau_{\mathrm{end}} = 10^{-5}$.
At the end of SU, the average Schmidt weight change is $\frac{1}{N}\sum_{n=1}^{N} ||\lambda_n(t+\tau) - \lambda_n(t)|| \lesssim 10^{-9}$, where $N$ is the total number of weights and $\lambda(t)$ is normalized.

During SU, we may control Schmidt weights to obtain appropriate uniform states.
After each imaginary time evolution step, we can average all weights to get the $d+id$-wave uniform state, or we can average two weights along the $a$ and $c$ directions to get the $d_{xy}$ or $d_{x^2-y^2}$-wave uniform states.
To make this process more stable with fermion parity symmetry, we choose $D_{{\rm even}} = D_{{\rm odd}}$, where $D = D_{{\rm even}} + D_{{\rm odd}}$ and $D_{{\rm even}}$ ($D_{{\rm odd}}$) denotes the size of the vector space with even (odd) parity.

\section{Contraction Scheme}
\label{sec:CS}

We use the variational uniform matrix product state (VUMPS) method \cite{10.21468/SciPostPhysLectNotes.7,PhysRevB.97.045145,PhysRevB.98.235148} to contract the 2D tensor network, which is a boundary MPS method working in the thermodynamic limit.
As shown in Fig. \ref{fig:19}(a, c), the scheme will be implemented in two steps. 
The first is contraction along the vertical direction ($\vec{v}_2$) to obtain the boundary MPS $A^U_{x,y}$ and $A^D_{x,y}$, and the second is contraction along the horizontal direction ($\vec{v}_1$) to obtain the left and right fixed points $E^L_{x,y}$ and $E^R_{x,y}$.
Bipartite sites on the honeycomb lattice are treated as a single site on the square lattice by defining $\mathbb{T}_{x,y} = \mathcal{C}_v\left(\mathcal{T}_{xy,A}\mathcal{T}_{xy,B}\bar{\mathcal{T}}_{xy,A}\bar{\mathcal{T}}_{xy,B}\right)$ in Fig. \ref{fig:19}(b). 
By swapping the horizontal and vertical directions, we can calculate two more boundary MPS $A^L_{x,y}$ and $A^R_{x,y}$, as well as two more fixed points $E^U_{x,y}$ and $E^D_{x,y}$ along the vertical direction.
    
As shown in Fig. \ref{fig:19}(d), the main process of the VUMPS algorithm is to solve a fixed point equation for the boundary MPS with $L_1\times L_2$ tensors $\mathbb{T}_{x,y}$ in the form of MPO, where $x=1,\cdots ,L_1$ and $y=1,\cdots ,L_2$.
We represent $A^U_{x,y}$ as a mixed canonical form $A^{U}_{x,y} = \{ C^U_{x,y}, A^{U,C}_{x,y}, A^{U,L}_{x,y}, A^{U,R}_{x,y}\}$, where $A^{U,C}_{x,y} = A^{U,L}_{x,y} C^U_{x,y} = C^U_{x-1,y} A^{U,R}_{x,y}$ and it satisfies the orthogonality condition $A^{U,L}_{x,y}\bar{A}^{U,L}_{x,y} = \mathbb{I}$ and $A^{U,R}_{x,y}\bar{A}^{U,R}_{x,y} = \mathbb{I}$.
This problem can be solved by variationally maximizing the overlap
\begin{equation}
\max_A \frac{|\langle \Psi(\bar{A})|\Psi(A')\rangle|^2}{\langle\Psi(\bar{A})|\Psi(A)\rangle}.
\end{equation}
Here, $|\Psi(A')\rangle$ refers to the combined MPO-MPS state $\mathcal{C}_v( A^{U,L}_{x-1,y} A^{U,C}_{x,y} A^{U,R}_{x+1,y} \cdots \mathbb{T}_{x-1,y} \mathbb{T}_{x,y} \mathbb{T}_{x+1,y} \cdots)$, and $|\Psi(A)\rangle$ refers to $\mathcal{C}_v( A^{U,L}_{x-1,y-1} A^{U,C}_{x,y-1} A^{U,R}_{x+1,y-1} \cdots)$.
    
The procedure for optimizing the overlap is illustrated in Fig. \ref{fig:19}(e).
First, the left fixed point $F^L_{x,y}$ is calculated by 
\begin{equation}
F^L_{L_1,y} \propto F^L_{L_1,y} \prod_{x=1}^{L_1} M_{x,y},
\end{equation}
where $M_{x,y} = A^{U,L}_{x,y}\mathbb{T}_{x,y} \bar{A}_{x,y-1}^{U,L}$ and $F^L_{x+1,y} = F^L_{x,y} M_{x+1,y}$. 
The right fixed point $F^R_{x,y}$ is obtained in the same way.
Next, new tensors $\tilde{A}^{U,C}_{x,y-1}$ and $\tilde{C}^U_{x,y-1}$ are updated by 
\begin{align}
\tilde{A}^{U,C}_{x,y-1} &= \lambda_A A^{U,C}_{x,y}F^L_{x-1,y}\mathbb{T}_{x,y}F^R_{x+1,y}, \\
\tilde{C}^{U}_{x,y-1} &= \lambda_C C^{U}_{x,y}F^L_{x,y}F^R_{x+1,y}.
\end{align}
After that, new $\tilde{A}^{U,L}_{x,y-1}$ and $\tilde{A}^{U,R}_{x,y-1}$ are obtained by solving equations $\epsilon_L = \min \| \tilde{A}^{U,C}_{x,y-1} - \tilde{A}^{U,L}_{x,y-1} \tilde{C}^U_{x,y-1} \|$ and $\epsilon_R = \min \| \tilde{A}^{U,C}_{x,y-1} - \tilde{C}^U_{x-1,y-1}\tilde{A}^{U,R}_{x,y-1} \|$.
Using polar decompositions \cite{10.21468/SciPostPhysLectNotes.7,PhysRevB.97.045145}, we have 
\begin{align}
\tilde{A}_{x,y-1}^{U,C} &= U_{{x,y-1}}^{\tilde{A},l} P_{{x,y-1}}^{\tilde{A},l} \quad \tilde{C}_{x,y-1}^U = U_{{x,y-1}}^{\tilde{C},l} P_{{x,y-1}}^{\tilde{C},l},\\
\tilde{A}_{x,y-1}^{U,C} &= P_{{x,y-1}}^{\tilde{A},r} U_{{x,y-1}}^{\tilde{A},r} \quad \tilde{C}_{x,y-1}^U = P_{{x,y-1}}^{\tilde{C},r} U_{{x,y-1}}^{\tilde{C},r}.
\end{align}
The left and right canonical forms are given by
\begin{align}
\tilde{A}^{U,L}_{x,y-1} &= U_{{x,y-1}}^{\tilde{A},l} (U_{{x,y-1}}^{\tilde{C},l})^\dagger, \\
\tilde{A}^{U,R}_{x,y-1} &= (U_{{x-1,y-1}}^{\tilde{C},r})^\dagger U_{{x,y-1}}^{\tilde{A},r}.
\end{align}
The above three steps are repeated until the gradient norms  $\epsilon_L$ and $\epsilon_R$ converge.
Following that, we go to the next row and update the new $A^U_{x,y-2}$.
The boundary MPS $A^D_{x,y}$ in the down direction is computed in the same way.

In the next step, we contract the tensor network shown in Fig. \ref{fig:19}(c).
The left and right effective environments $E^L_{x,y}$, $E^R_{x,y}$ are the fixed points of the transfer matrix $\prod_{x=1}^{L_1} T_{x,y}$, where $T_{x,y}$ is constructed by $A^{U}_{x,y} \mathbb{T}_{x,y} A^D_{x,y}$.

The boundary MPS and fixed points along the vertical direction are computed in a similar way as in the horizontal direction.
Physical quantities can be calculated using the environmental tensors $\{A^\alpha_{x,y}, E^\alpha_{x,y}\}$, where $\alpha = U, D, L, R$ denotes the four directions.
The contraction method has an overall complexity of $\mathcal{O}(\chi^2 D^5$).

\bibliography{refs}

\end{document}